# Massive feature extraction for explaining and foretelling hydroclimatic time series forecastability at the global scale

Georgia Papacharalampous[1,*], Hristos Tyralis[2,3], Ilias G. Pechlivanidis[4], Salvatore Grimaldi[5,6], and Elena Volpi[1]

[1] Department of Engineering, Roma Tre University, Via V. Volterra 62, 00146 Rome, Italy

[2] Department of Water Resources and Environmental Engineering, School of Civil Engineering, National Technical University of Athens, Heroon Polytechneiou 5, 15780 Zographou, Greece

[3] Hellenic Air Force General Staff, Hellenic Air Force, Mesogion Avenue 227–231, 15561 Cholargos, Greece

[4] Swedish Meteorological and Hydrological Institute, 60176 Norrköping, Sweden

[5] Department for Innovation in Biological, Agro-food and Forest Systems, University of Tuscia, Viterbo, Italy

[6] Department of Mechanical and Aerospace Engineering, Tandon School of Engineering, New York University, Brooklyn, NY, 10003, USA

* Correspondence: papacharalampous.georgia@gmail.com, tel: +30 69474 98589

**Email addresses and ORCID profiles:** papacharalampous.georgia@gmail.com, https://orcid.org/0000-0001-5446-954X (Georgia Papacharalampous); montchrister@gmail.com, hristos@itia.ntua.gr https://orcid.org/0000-0002-8932-4997 (Hristos Tyralis); ilias.pechlivanidis@gmail.com; https://orcid.org/0000-0002-3416-317X (Ilias Pechlivanidis); salvatore.grimaldi@unitus.it, http://orcid.org/0000-0001-5715-106X (Salvatore Grimaldi); elena.volpi@uniroma3.it, https://orcid.org/0000-0002-9511-1496 (Elena Volpi)

**Abstract:** Statistical analyses and descriptive characterizations are sometimes assumed to be offering information on time series forecastability. Despite the scientific interest suggested by such assumptions, the relationships between descriptive time series features (e.g., temporal dependence, entropy, seasonality, trend and linearity features) and actual time series forecastability (quantified by issuing and assessing forecasts for the past) are scarcely studied and quantified in the literature. In this work, we aim to fill in this gap by investigating such relationships, and the way that they can be exploited for understanding hydroclimatic forecastability and its patterns. To this end, we follow a systematic framework bringing together a variety of –mostly new for hydrology– concepts and methods, including 57 descriptive features and nine seasonal time series forecasting methods (i.e., one simple, five exponential smoothing, two state space and one automated autoregressive fractionally integrated moving average methods). We apply this framework to three global datasets originating from the larger Global Historical Climatology Network (GHCN) and Global Streamflow Indices and Metadata (GSIM) archives. As these datasets comprise over 13 000 monthly temperature, precipitation and river flow time series from several continents and hydroclimatic regimes, they allow us to provide trustable characterizations and interpretations of 12-month ahead hydroclimatic forecastability at the global scale. We first find that the exponential smoothing and state space methods for time series forecasting are rather equally efficient in identifying an upper limit of this forecastability in terms of Nash-Sutcliffe efficiency, while the simple method is shown to be mostly useful in identifying its lower limit. We then demonstrate that the assessed forecastability is strongly related to several descriptive features, including seasonality, entropy, (partial) autocorrelation, stability, (non)linearity, spikiness and heterogeneity features, among others. We further (i) show that, if such descriptive information is available for a monthly hydroclimatic time series, we can even foretell the quality of its future forecasts with a considerable degree of confidence, and (ii) rank the features according to their efficiency in explaining and foretelling forecastability. We believe that the obtained rankings are of key importance for understanding forecastability. Spatial forecastability patterns are also revealed through our experiments, with East Asia (Europe) being characterized by larger (smaller) monthly temperature time series forecastability and the Indian subcontinent (Australia) being characterized by larger (smaller) monthly precipitation time series forecastability, compared to other continental-scale regions, and less notable differences characterizing monthly river flow from continent to continent. A comprehensive interpretation of such patters through massive feature extraction and feature-based time series clustering is shown to be possible. Indeed, continental-scale regions characterized by different degrees of forecastability are also attributed to different clusters or mixtures of clusters (because of their essential differences in terms of descriptive features).

1.    Introduction

Large-sample (e.g., global-scale) hydrological studies (e.g., Fang et al. 2018; Markonis et al. 2018a,b; Papacharalampous et al. 2018; Blöschl et al. 2019b; Ceola et al. 2019; Papalexiou and Montanari 2019; Tyralis et al. 2019b; Pechlivanidis et al. 2020; Girons Lopez et al. 2021; Messager et al. 2021) are conducted with increasing frequency nowadays, as computer power is rapidly increasing, numerous open-source packages are released (e.g., in the R and Python programming languages, documented in R Core Team 2021 and Python Software Foundation 2021, respectively) and several large hydrological datasets are made publicly available (see, e.g., the documentations in Newman et al. 2015; Addor et al. 2017; Alvarez-Garreton et al. 2018; Do et al. 2018; Gudmundsson et al. 2018; Menne et al. 2018; Althoff et al. 2020; Chagas et al. 2020; Coxon et al. 2020; Fowler et al. 2021; Klingler et al. 2021). Such studies are indeed necessary (Gupta et al. 2014) both for creating new earth system science knowledge, and for improving modelling frameworks in hydrology. The complementary importance of these two major targets is recognised within initiatives (Montanari et al. 2013; Blöschl et al. 2019a), and largely drives scientific curiosity and practical considerations in the various hydrological sub-disciplines.

Addressing each of these complementary targets requires both investigations conducted from the descriptive perspective and investigations conducted from the predictive (or predictability) perspective (e.g., Papacharalampous and Tyralis 2020). The largest part of statistical hydrology, and perhaps even more the largest part of the large-sample studies in this sub-discipline, contribute to the total research efforts in the field by performing investigations of the former type (see, e.g., the works by Scheidegger 1970; Yevjevich 1987; Montanari 2003; Markonis and Koutsoyiannis 2013; Koutsoyiannis 2013; Ledvinka 2015; Ledvinka and Lamacova 2015; Volpi 2019; Serinaldi et al. 2020;

Papacharalampous et al. 2021; see also the overview in Grimaldi et al. 2011). On the other hand, the largest part of the various hydrological prediction methods and tests (see, e.g., those by Xu et al. 2018; Curceac et al. 2019, 2020; Khatami et al. 2019, 2020; Quilty et al. 2019; Quilty and Adamowski 2020; Pechlivanidis et al. 2020; Rahman et al. 2020; Althoff et al. 2021; Sikorska-Senoner and Quilty 2021; Széles et al. 2021; see also the overview by Sivakumar and Berndtsson 2010) can be (mostly) attributed to other hydrological sub-disciplines. The latter sub-disciplines include (but are not limited to) those few sub-disciplines that are fully devoted to prediction (e.g., the various forecasting sub-disciplines), with the time series analysis and clustering inputs to most of their methods being minimal (if not absent) currently. They also include the catchment hydrology sub-discipline, in which catchment (e.g., geologic, topographic, soil and land use) attributes, time series analysis, clustering and prediction concepts are regularly merged and integrated into broader methodological frameworks (see, e.g., the works by Pechlivanidis and Arheimer 2015; Széles et al. 2018; Iliopoulou et al. 2019; Tyralis et al. 2019b; Knoben et al. 2020; Sikorska-Senoner and Seibert 2020).

Such integrations can offer valuable inspirations and can progress our understanding of the interplay between analysis and prediction, as well as the connections between improved earth system knowledge and improved modelling frameworks. More generally, the culture of merging concepts and methods from various fields and from various schools of thought (see, e.g., the discussions in Breiman 2001b; Shmueli 2010) could lead us to new fruitful avenues (see the related discussions in Todini 2007; Moallemi et al. 2021; Sidle 2021). Recognizing open opportunities sometimes requires a fresh look at old pathways, traditions and viewpoints, which can be supported simply by an openness to various concepts and methods. For instance, investigations on hydrometeorological and hydroclimatic time series forecasting are sometimes perceived to predominantly (if not solely) support practical considerations and needs (and, therefore, they are conducted accordingly most of the times), although they can also lead to significant advancements in terms of earth system understanding through time series forecastability assessments and comparisons across regions (and climates), and/or investigations of the relationships between time series forecastability and descriptive time series features (Papacharalampous and Tyralis 2020). We define time series forecastability as a "*time series feature for measuring the degree in which the future of a time series can be predicted based exclusively on its past*".

Interesting discussions on hydrometeorological and hydroclimatic time series forecastability sometimes accompany authentic experiments (e.g., those by Dimitriadis et al. 2016), or hydrometeorological and hydroclimatic time series analyses (e.g., in Koutsoyiannis 2010, 2013; Dimitriadis et al. 2021) in the statistical hydrology literature. In agreement with literature pieces from other fields (see, e.g., Goerg 2013), (some of) these discussions naturally assume that time series forecastability can be inferred by computing descriptive time series features of specific types, such as entropy features (see, e.g., Goerg 2013) and the Hurst parameter of the fractional Gaussian noise process (see, e.g., Mandelbrot and Wallis 1968; Beran 1994), and indicate considerable scientific and practical interest in the overall topic. Despite this indisputable interest, the relationships between descriptive time series features (e.g., temporal dependence, entropy, seasonality and trend features) and actual time series forecastability (in principle, quantified by issuing and assessing forecasts for the past, with the observed values corresponding to these forecasts being treated as unknown and unseen at the time of the forecast; see again the definition) are scarcely studied and quantified in the literature, although the means for conducting the required large-sample and massive investigations are available today.

In this work, we aim to fill in this gap by extensively merging –for the first time in the literature– various concepts and methods from the statistical-stochastic hydrology (see Papacharalampous and Tyralis 2020; Papacharalampous et al. 2021), time series forecasting (see, e.g., Armstrong 2001; De Gooijer and Hyndman 2006; Hyndman and Athanasopoulos 2021; Kang et al. 2017; Talagala et al. 2019; Ponce-Flores et al. 2020), data science (see, e.g., Fulcher et al. 2013; Hyndman et al. 2015; Donoho 2017; Fulcher and Jones 2017) and machine learning (see, e.g., Hastie et al. 2009; Alpaydin 2010; James et al. 2013) fields, as well as concepts from other hydrological and geoscientific sub-disciplines (see Knoben et al. 2020; Pechlivanidis et al. 2020; Girons Lopez et al. 2021; Manero Font and Béjar Alonso 2021), into a massive and detailed framework for quantifying the relationships between hydrometeorological time series forecastability and descriptive time series features, and by applying this framework to three global datasets. The latter comprise over 13 000 monthly temperature, precipitation and river flow time series. These time series have been recorded in stations covering a large part of the earth's surface; thus, they allow us to provide trustable characterizations and interpretations of 12-month ahead forecastability (i.e., forecastability of the next 12 months of a monthly time series; Taieb et al. 2012) of temperature, precipitation and river flow time series at the global scale. Overall, we aspire to contribute to the field of statistical and comparative hydrology by providing trustable answers to the following research questions:

- To what extent can descriptive time series features be exploited for inferring and foretelling monthly temperature, precipitation and river flow time series forecastability?
- What are the rankings of various descriptive features according to the information that they offer for fulfilling the above task?
- Which locations around the globe and which climates are characterized by the largest or smallest 12-month ahead forecastability based on their temperature, precipitation and river flow time series records, and to what extent can we interpret this knowledge by computing descriptive features and by performing feature-based time series clustering?

## 2. Data

To answer the above scientific questions, we use three open global datasets that not only describe key hydroclimatic variables, i.e., temperature, precipitation and river flow, but also provide long time series records to perform a robust statistical investigation. These datasets are subsets of larger freely available ones (Do et al. 2018; Menne et al. 2018; Peterson and Vose 1997; see also Appendix A), and comprise 2 432 mean monthly temperature time series, 5 071 total monthly precipitation time series and 5 601 mean monthly river flow time series. These time series are 40-year-long, starting in January and ending in December. They originate from stations that are sufficiently scattered around the globe, with North America and Europe being well represented in all the three datasets, Asia being well represented in the temperature and precipitation datasets, and the remaining continents being well represented in the precipitation dataset and not absent from the temperature and precipitation datasets. The same 13 104 monthly time series have been first aggregated and utilized into a single framework in the work by Papacharalampous et al. (2021), where more details on their selection are available. In summary, the temperature and precipitation time series are quality assured already from their original formation (Menne et al. 2018; Peterson and Vose 1997), while the river flow time series are not contaminated by irregularities that might be due to human activities (e.g., by abrupt changes in their mean and variance), as assured through visual quality inspections (Papacharalampous et al. 2021).

Additionally to the above-outlined monthly hydroclimatic time series data, we use gridded climate data (Kottek et al. 2006; see also Appendix A). Based on this latter data, we assign the temperature, precipitation and river flow stations at which the selected time series have been recorded to their main climate class (i.e., the equatorial, arid, warm temperate, snow and polar climates). Only polar climates are found to be under-represented (with less than 30 time series records) in the monthly precipitation and river flow time series datasets.

## 3. Methodology

### 3.1 Time series forecasting methods

Our methodological framework adopts, among others, nine mostly parsimonious and well-established and/or well-tested time series forecasting methods from the forecasting field. These methods have been found to outperform others in forecasting competitions (see, e.g., Fildes 2020; Hyndman 2020); therefore, they are selected in this work for collectively supporting monthly time series forecastability characterizations. Their parameters and full documentations can be found in specialized textbooks and articles, and are here omitted for reasons of brevity, in line with the guidelines by Abrahart et al. (2008). In brief, the selected time series forecasting methods are the:

o  Seasonal benchmark, which sets the time series forecast for each month equal to the last year's observation for the same month in a 12-month ahead forecasting setting (i.e., in a setting where we forecast the future 12 months of a monthly time series; see again Taieb et al. 2012). It is usually referred to as the "persistent" method in the forecasting field (see also its non-seasonal version, e.g., in Papacharalampous et al. 2019), and as "persistency" in the hydrological forecasting field. In the latter field, it is commonly used in forecasting short to medium ranges.

o  Seasonal exponential smoothing (ETS) without trend, which is the simplest available model from the exponential smoothing family (see, e.g., Brown 1959; Winters 1960; Hyndman et al. 2002; Holt 2004; Hyndman et al. 2008a; Hyndman et al. 2008b) according to the classification by Taylor (2003) with the ability to model seasonality (Gardner 2006). For this latter purpose, it can perform either additive or multiplicative time series decomposition. Here, the former option is adopted. Also, the parameters of the method are optimized through minimization of the mean square error.

- Seasonal ETS with non-damped trend, which is an extension of the right above time series forecasting method being able to model trends (see, e.g., Taylor 2003).
- Seasonal ETS with damped trend, which is an extension of the two right above time series forecasting methods incorporating trends that are damped as the horizon increases (see, e.g., Taylor 2003).
- Seasonal case-informed ETS, which automatically performs an optimal selection among the three right above models by computing the Akaike Information Criterion (AIC; Akaike 1974). This method has been proven accurate on data from the M3-competition (see Hyndman et al. 2002).
- Seasonal autoregressive integrated moving average (ARIMA; Box and Jenkins 1970), which constitutes an extension of ARIMA models that is able to model seasonality (see, e.g., Hipel and McLeod 1994, Chapter 12; Wei 2006, Chapter 8.3), and has been further modified for supporting automatic time series forecasting applications (see, e.g., Hyndman and Athanasopoulos 2021, Chapters 8.6). It computes the AIC for selecting an optimum model (by also taking into consideration predefined restrictions), and mixes the maximum likelihood method with a method minimizing the sum of squared residuals for estimating the model's parameters.
- Seasonal TBATS, which is a trigonometric exponential smoothing state space method with Box-Cox transformation, autoregressive moving average (ARMA) errors, trend and seasonal components by De Livera et al. (2011). This method has the ability to model complex seasonal patterns.
- Seasonal BATS, which is an exponential smoothing state space method with Box-Cox transformation, ARMA errors, trend and seasonal components (see its detailed description, e.g., in De Livera et al. 2011).
- Seasonal complex ETS, i.e., the seasonal version of the non-linear time series forecasting method by Svetunkov and Kourentzes (2016). This method is based on the theory of complex variable functions and does not perform time series decomposition (in contrast to most exponential smoothing methods). It performs optimal model selection by computing the AIC and estimates its parameters by minimizing the mean square error.

For characterizing hydroclimatic time series forecastability, the performances of the above-outlined time series forecasting methods are assessed and further summarized under the notions of the "worst" and "best" methods for each station, as detailed in Section 3.2. Under these notions, we examine 11 time series forecasting methods.

**3.2 Forecast evaluation setup and descriptive time series features**

We apply the above time series forecasting methods (for the utilized statistical software, see the Supplementary Data, Text S1; Appendix B) to the three time series datasets (see Section 2). Specifically, we issue 12-month ahead forecasts (see again their definition above) corresponding to the last 10 years of the time series. This is made separately for each time series according to the following procedures: (a) first, we fit the models to the first 30 years and forecast the 31st year; (b) we subsequently refit the previously fitted models to the first 31 years and forecast the 32nd year; and (c) so on until we get the entire 10-year long time series of 12-month ahead forecasts. For assessing the forecast quality in relative terms, we compute the Nash-Sutcliffe efficiency (Nash and Sutcliffe 1970) of the 10-year-long 12-month ahead forecasts according to Supplementary Data (Text S2). We have here selected this specific metric because of its popularity in hydrology in general and in hydrological forecasting in particular (see, e.g., the review by Cheng et al. 2017), as well as its considerable degree of interpretability. In fact, this metric takes values between $-\infty$ and 1 (perfect forecast), with values larger (smaller) than 0 indicating more (less) accurate forecasts than the mean of the true values corresponding to these forecasts. Furthermore, we believe that the length of the testing period (i.e., 120 time series points) is sufficient, given also the large number of the examined time series. Based on the computed Nash-Sutcliffe efficiency values, we identify the worst and best methods for each station, and summarize their forecast quality (as if they were the tenth and eleventh time series forecasting methods of this work).

We further compute 57 largely diverse descriptive time series features (Hyndman et al. 2020; see also Fulcher et al. 2013; Hyndman et al. 2015; Fulcher and Jones 2017; Kang et al. 2017, 2020). These features are presented in the Supplementary Data (Text S3), and have already been computed and proposed as a set for massive feature extraction in statistical hydrology and environmental science (with small deviations with respect to this work, as detailed in the Supplementary Data, Text S3) by Papacharalampous et al. (2021). Two additional features appearing in this latter study (i.e., the

`localsimple_mean1` and `localsimple_lfitac` ones) are not considered herein, as their computation involves prediction experiments. Indeed, this makes them irrelevant to the problem of explaining hydroclimatic time series forecastability. We compute each of the 57 features twice; the first time for the entire 40-year-long time series and the second time for their eldest 30-year-long segments (thereby leaving out their most recent 10-year-long segments, which correspond to the same time periods as the delivered forecasts). We, therefore, form two feature datasets for each time series type, i.e., six feature datasets in total. The total number of the computed feature values is 13 104 (number of time series) × 57 (number of descriptive features) × 2 (number of times that each feature is computed for each time series) = 1 493 856.

### 3.3 Time series forecastability comparisons across clusters

Based on the descriptive features of the entire 40-year-long time series (i.e., based on three of the six feature datasets; see right above), we divide the datasets into clusters by using unsupervised random forests (for time series clustering; see, e.g., Yan et al. 2013) with 5 000 trees. The related procedures compose a feature-based time series clustering methodology, which constitutes a close variant of the original one by Papacharalampous et al. (2021). We get five temperature, five precipitation and five river flow time series clusters. The selected number of clusters is sufficient for supporting our aims here; however, it is also indicative, as a different (e.g., larger) number of clusters would also be sufficient. We use the delivered clusters to examine the existence of possible spatial patterns of monthly temperature, precipitation and river flow time series forecastability at the global scale. The time series forecastability comparisons across clusters are summarized by drawing side-by-side boxplots and by reporting median-case Nash-Sutcliffe efficiency values.

### 3.4 Time series forecastability comparisons across regions and climates

We form continental-scale regional groups of temperature, precipitation and river flow stations. These groups are not exhaustive; yet, they correspond to the regions with the highest density in stations. They are used, together with the delivered temperature, precipitation and river flow clusters (see Section 3.3), for investigating the existence of possible spatial patterns of monthly temperature, precipitation and river flow time series forecastability at the global scale. For each regional group of stations, we compute the percentage of the stations assigned to each cluster, and summarize its time series forecastability by drawing side-by-side boxplots of Nash-Sutcliffe efficiency values and by reporting the medians of these values. The key idea behind these procedures is to allow in parallel explorations across clusters and regional groups of stations, as these specific explorations could serve towards a better perception of monthly hydroclimatic time series forecastability at the global scale.

Moreover, we examine monthly temperature, precipitation and river flow time series forecastability in light of the additional information offered by the gridded climate data (see Section 2). Similar to the time series forecastability comparisons across clusters and regional groups, the time series forecastability comparisons across main climatic types are summarized by drawing side-by-side boxplots and by reporting median-case Nash-Sutcliffe efficiency values.

### 3.5 Time series forecastability versus descriptive time series features

For all the six feature datasets (see Section 3.2), we examine the relationships between time series forecastability in terms of the Nash-Sutcliffe efficiency (which is assumed to be well represented by the Nash-Sutcliffe efficiency of the best method; see Section 3.2) and the various time series features, i.e., 6 (number of feature datasets) × 57 (number of descriptive features) = 342 relationships. The related experiments involve two independent procedures, with the latter being the most important one (as it allows an objective ranking of the features as regards their contribution in interpreting or foretelling monthly temperature, precipitation and river flow time series forecastability): We (i) compute Spearman correlations (Spearman 1904) to characterize the relationships with respect to their intensity, and (ii) apply random forests (Breiman 2001a; see also the review by Tyralis et al. 2019a) with 500 trees to assess the relative importance of the descriptive features in predicting the Nash-Sutcliffe efficiency values, and rank the descriptive features based on this importance. Random forests are indeed among the ideal machine learning algorithms for this task (see also the relevant discussions in Tyralis and Papacharalampous 2021).

The above-outlined procedures (i) and (ii) are repeated for each of the temperature, precipitation and river flow clusters of the descriptive features of the entire 40-year-long time series (i.e., for three of the six feature datasets; see Section 3.2), i.e., for 3 (number of corresponding feature datasets) × 5 (number of clusters per feature dataset) × 57 (number of descriptive features) = 855 additional relationships (with the focus being retained on the best method). Lastly, procedure (ii) is repeated for the relationships between the descriptive features of the entire 40-year-long time series and the Nash-Sutcliffe efficiency offered by each of the remaining time series forecasting methods (including the worst method), i.e., for 3 (number of corresponding feature datasets) × 57 (number of descriptive features) × 10 (number of the remaining time series forecasting methods) = 1 710 additional relationships.

## 4. Results

### 4.1 Diagnosis of time series forecastability over clusters and regional groups

The temperature, precipitation and river flow clusters (C-T1, C-T2, C-T3, C-T4, C-T5, C-P1, C-P2, C-P3, C-P4, C-P5, C-R1, C-R2, C-R3, C-R4 and C-R5), obtained by using the features of the entire 40-year-long monthly temperature, precipitation and river flow time series, hold an important position throughout this paper. Information about these clusters is presented in Figure 1, and is examined together with the forecasting performance of nine time series methods in terms of Nash-Sutcliffe efficiency (see Figure 2) to further reveal the considerable extent to which actual time series forecastability in terms of Nash-Sutcliffe efficiency is related to descriptive time series features for monthly temperature, precipitation and river flow. Moreover, it allows extensive comparisons of these hydrometeorological variables in terms of this knowledge, and reveals spatial patterns of their 12-month ahead time series forecastability at the global scale. At the same time, the large groups of temperature, precipitation and river flow stations (G-T1, G-T2, G-T3, G-P1, G-P2, G-P3, G-P4, G-P5, G-P6, G-R1, G-R2 and G-R3; see Figure 1) allow additional investigations of these patterns. These latter groups are compared with respect to their 12-month ahead time series forecastability in terms of Nash-Sutcliffe efficiency in Figure 3.

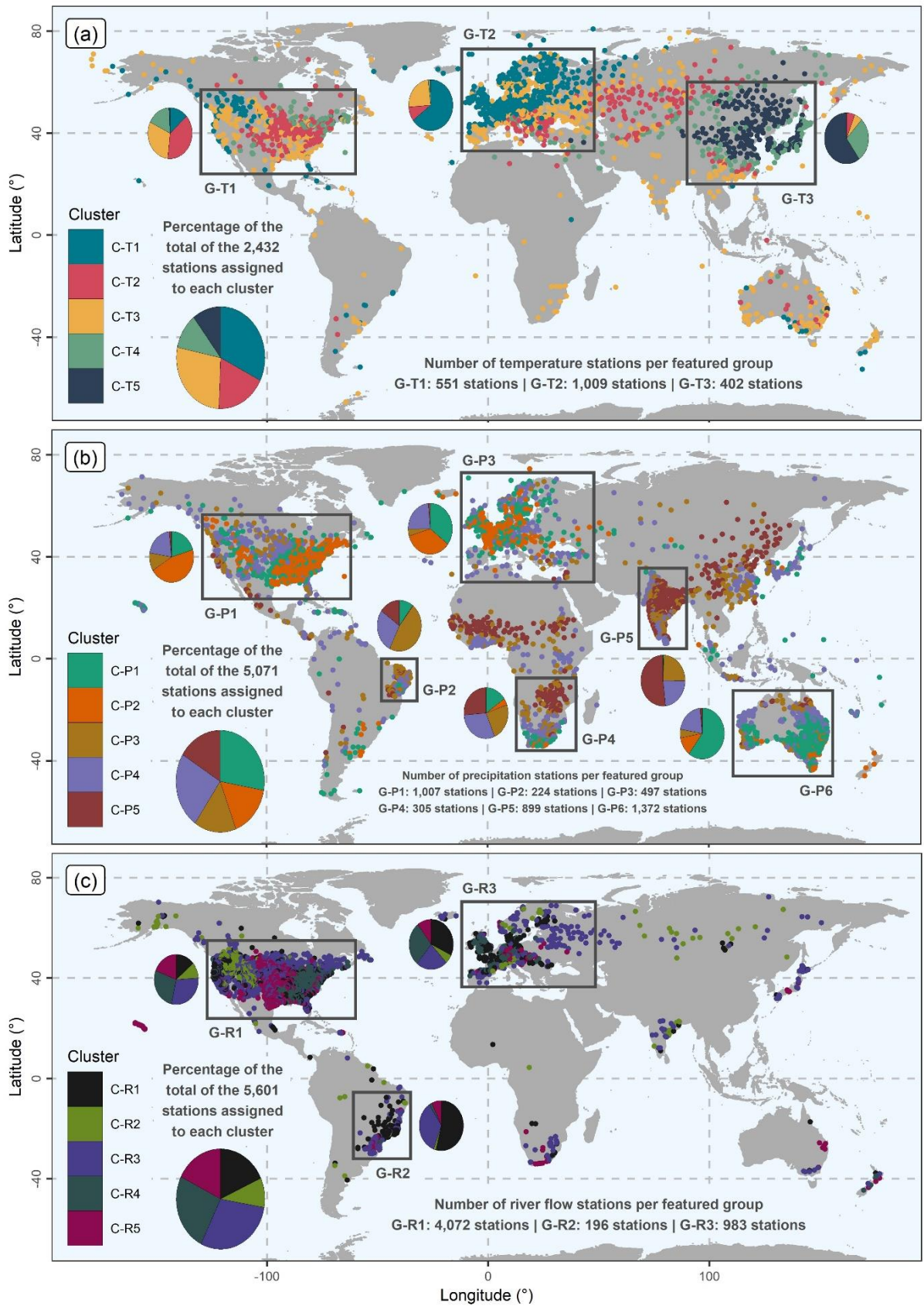

Figure 1. Geographical locations of the utilized (a) temperature, (b) precipitation and (c) river flow stations, definition of regional groups of stations, and summary of the clustering outcomes. The latter have been obtained using the entire 40-year-long time series.

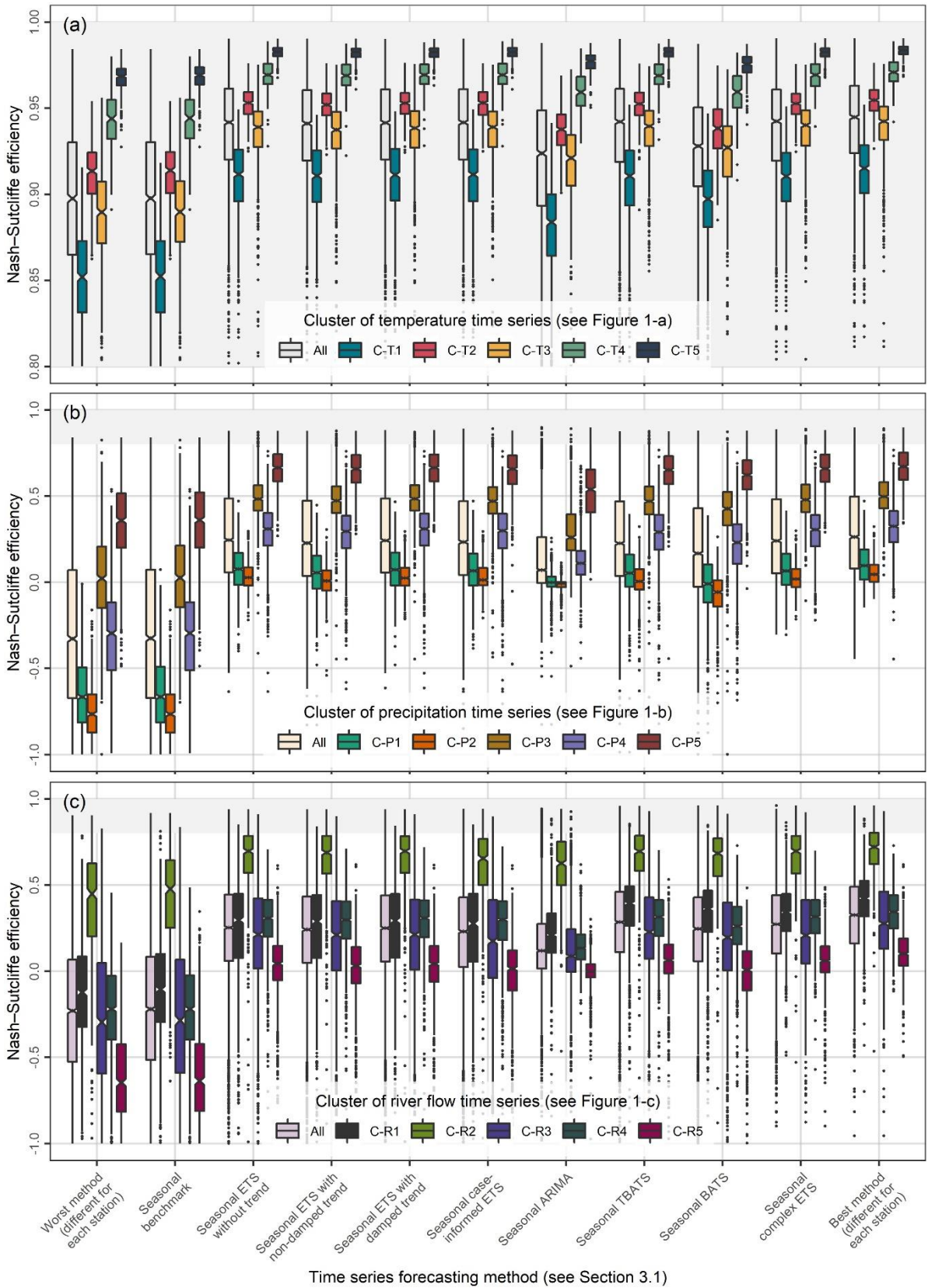

Figure 2. Summary of the results concerning the (a) temperature, (b) precipitation and (c) river flow time series forecastability assessment. The presentation is made per cluster. The clusters have been obtained using the 40-year-long time series. To allow comparisons across clusters, the vertical axes have been truncated at 0.80 in (a) and −1.00 in (b, c).

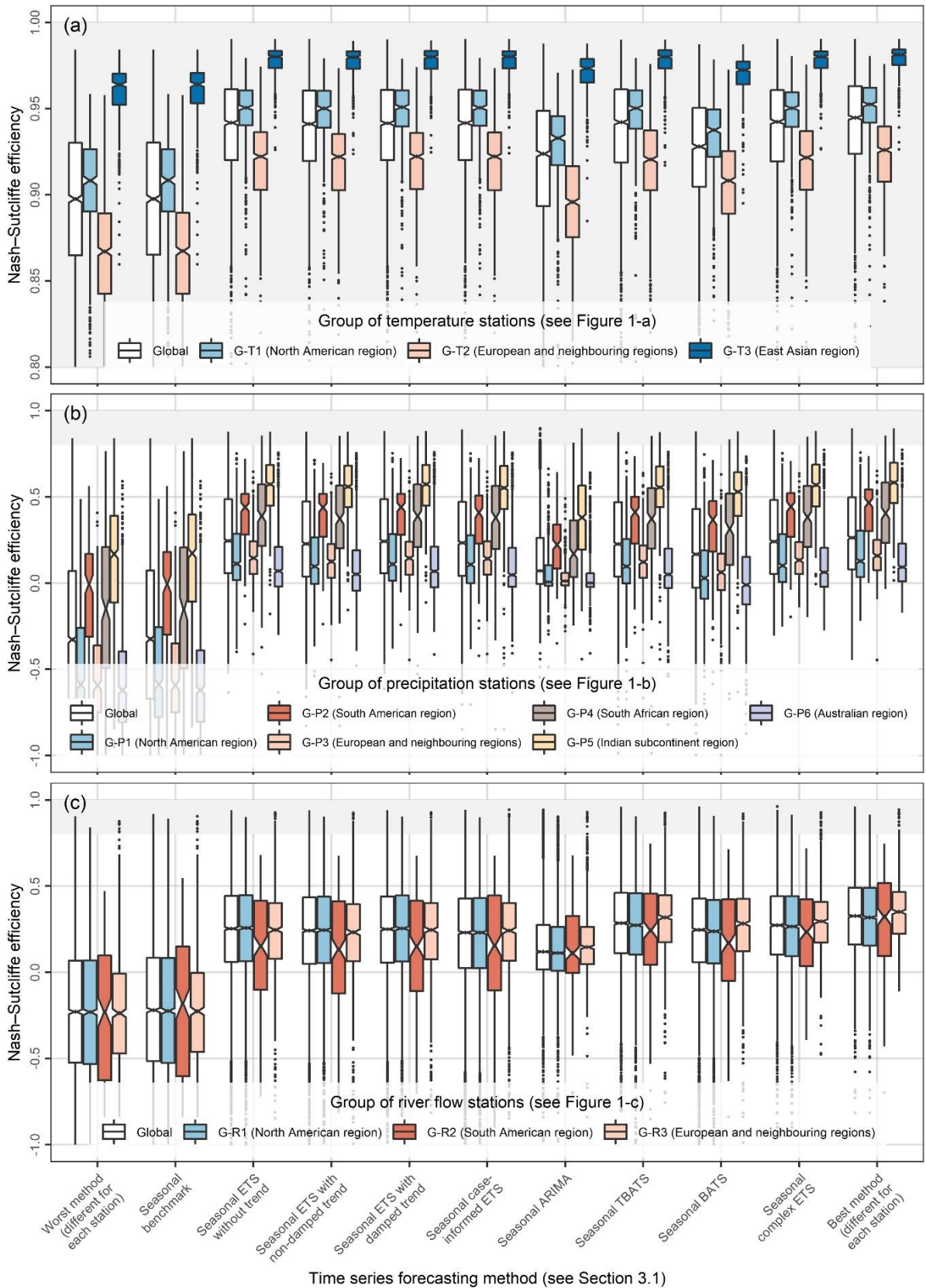

Figure 3. Summary of the results concerning the (a) temperature, (b) precipitation and (c) river flow time series forecastability assessment. The presentation is made per group of stations. Those groups of stations located in the same region are represented by the same colour independently of their type (i.e., temperature, precipitation or river flow).

The five temperature clusters are indeed characterized by different degrees of time series forecastability (see Figure 2a), and the same holds for the three groups of temperature stations covering different continental-scale regions (see Figure 3a). More precisely, cluster C-T5 (a cluster comprising only time series observed in East Asia) is characterized by the largest forecastability, followed by cluster C-T4 (which comprises time series observed in East Asia, North America and some other locations). As by far its largest part is covered by temperature clusters C-T5 and C-T4, East Asia (group G-T3) has the largest monthly temperature time series forecastability. On the other hand, the smallest forecastability is found to characterize the largest part of the monthly temperature time series observed in Europe, as well as Europe itself (group G-T2) with respect to its monthly temperature time series, and some other monthly temperature time series observed in North America and other locations (cluster C-T1). Moreover, clusters C-T2 and C-T3 and, thus, approximately half of the totally examined monthly temperature time series and the largest part of the monthly temperature time series from North America are characterized by medium forecastability. This latter information also explains why North America itself (group G-T1) is found to be characterized by a medium degree of forecastability with respect to its monthly temperature time series.

Different degrees of forecastability have been found to characterize the precipitation clusters (see Figure 2b) almost as much as they characterize the temperature clusters (see Figure 2a). However, not all of the precipitation station groups are characterized by notable differences in terms of forecastability (see Figure 3b). In detail, cluster C-P5 (a cluster mostly comprising time series from South America, Central and South Africa, the Indian subcontinent and East Asia) has been found to exhibit the largest forecastability. Larger forecastability than average is found to also characterize cluster C-P3, which comprises time series mostly from the same large regions as C-P5, while rather an average degree of forecastability characterizes cluster C-P4. This latter cluster comprises approximately the one fourth of the total number of time series in several different regions around the globe (see groups G-P1, G-P2, G-P3, G-P4, G-P5 and G-P6). Group G-P5 (a group located in the Indian subcontinent) is characterized by the largest monthly precipitation forecastability, as approximately half of the time series observed at its stations belong to cluster C-P5 and (most of) its remaining time series belong either to C-P3 or C-P4 with the respective percentages being approximately equal. Two other groups exhibiting larger monthly precipitation time series forecastability than average are G-P2 and G-P4 (groups located in South America and South Africa, respectively), as the larger parts of their time series are attributed mostly to clusters C-P3 and C-P4 but also to cluster C-P5. The largest part of the monthly precipitation time series with the smallest forecastability (attributed either to cluster C-P1 or to cluster C-P2) have been observed in North America, Europe or Australia.

The degree to which the various river flow clusters (see Figure 2c) are characterized by different magnitudes of time series forecastability is smaller than the respective degree characterizing the temperature and precipitation clusters (see Figure 2a,b). Two river flow clusters strongly deviate from the remaining three. The first one is cluster C-R2 (a cluster that mostly comprises time series observed in a North American region), which exhibits the largest forecastability, while the second one is C-R5. This latter cluster largely comprises monthly river flow time series observed in the central part of North America and exhibits the smallest forecastability. Despite the existence of these two clusters, station groups G-R1, G-R2 and G-R3 (and, therefore, their corresponding continental-scale regions) are characterized by a similar degree of forecastability, as most of the monthly river flow time series observed at their stations are assigned to one of the three clusters exhibiting medium forecastability (i.e., clusters C-R1, C-R3 and C-R4).

Figure 2c can be further discussed and interpreted from a catchment hydrology perspective, in light of existing works investigating the relationships between river flow predictability-forecastability and hydrological signatures (e.g., the base flow index, flashiness, slope of streamflow duration curve). In fact, river flow predictability-forecastability is strongly linked to the hydrological memory of the river system (Pechlivanidis et al. 2020; Harrigan et al. 2018; Girons Lopez et al. 2021). Hydrological regimes driven by seasonal snow processes (i.e., accumulation and melting; probably overlapping with parts of cluster C-R2) tend to have a persistent intra-annual pattern and are, hence, well forecasted. On the contrary, fast responding hydrological regimes (probably overlapping with parts of cluster C-R5) tend to strongly depend on the precipitation signal, which is subject to larger deviations in the seasonal distribution. Such regimes are characterized in detail in Pechlivanidis et al. (2020), while similar interpretations (from a process-oriented understanding viewpoint) for monthly temperature and precipitation time series forecastability could involve climatological pattern stability characterizations, among others. Extensive interpretations from a statistical hydrology perspective are provided herein, specifically in Section 4.4.

## 4.2 Diagnosis of time series forecastability over climatic types

Furthermore, we exploit the Nash-Sutcliffe efficiency values to compare the five main climate types (i.e., the equatorial, arid, warm temperate, snow and polar climates) with respect to their monthly temperature (see Figure 4a), precipitation (see Figure 4b) and river flow (see Figure 4c) time series forecastability. These main climates are defined by specific temperature or precipitation conditions (see Kottek et al. 2006, Table 1), which allow the growth of different vegetation groups. We find that the equatorial and arid climates are characterized by the smallest and largest monthly temperature time series forecastability, respectively. We also find that the equatorial climates are characterized by the largest monthly precipitation time series forecastability, and that the snow climates are characterized by the largest monthly river flow forecastability. We recall (from Section 2) that polar climates are under-represented in the monthly precipitation and river flow datasets; therefore, they are not included in the respective comparisons.

## 4.3 Summary of time series forecastability characterizations and comparisons

To further summarize the various time series forecastability characterizations discussed so far, in Figure 5 we present the medians of the side-by-side boxplots (Figures 2–4). We believe that these medians constitute a good measure of monthly temperature, precipitation and river flow time series forecastability, especially if we consider the popularity of the Nash-Sutcliffe efficiency in our field. Moreover, Figures 2–5 allow us to compare the forecasting methods with respect to their usefulness in assessing monthly hydroclimatic time series forecastability in terms of Nash-Sutcliffe efficiency. We observe that the seasonal exponential smoothing and state space methods are rather equally efficient in identifying an upper limit of this forecastability. We also observe that the seasonal benchmark is mostly useful in identifying its lower limit.

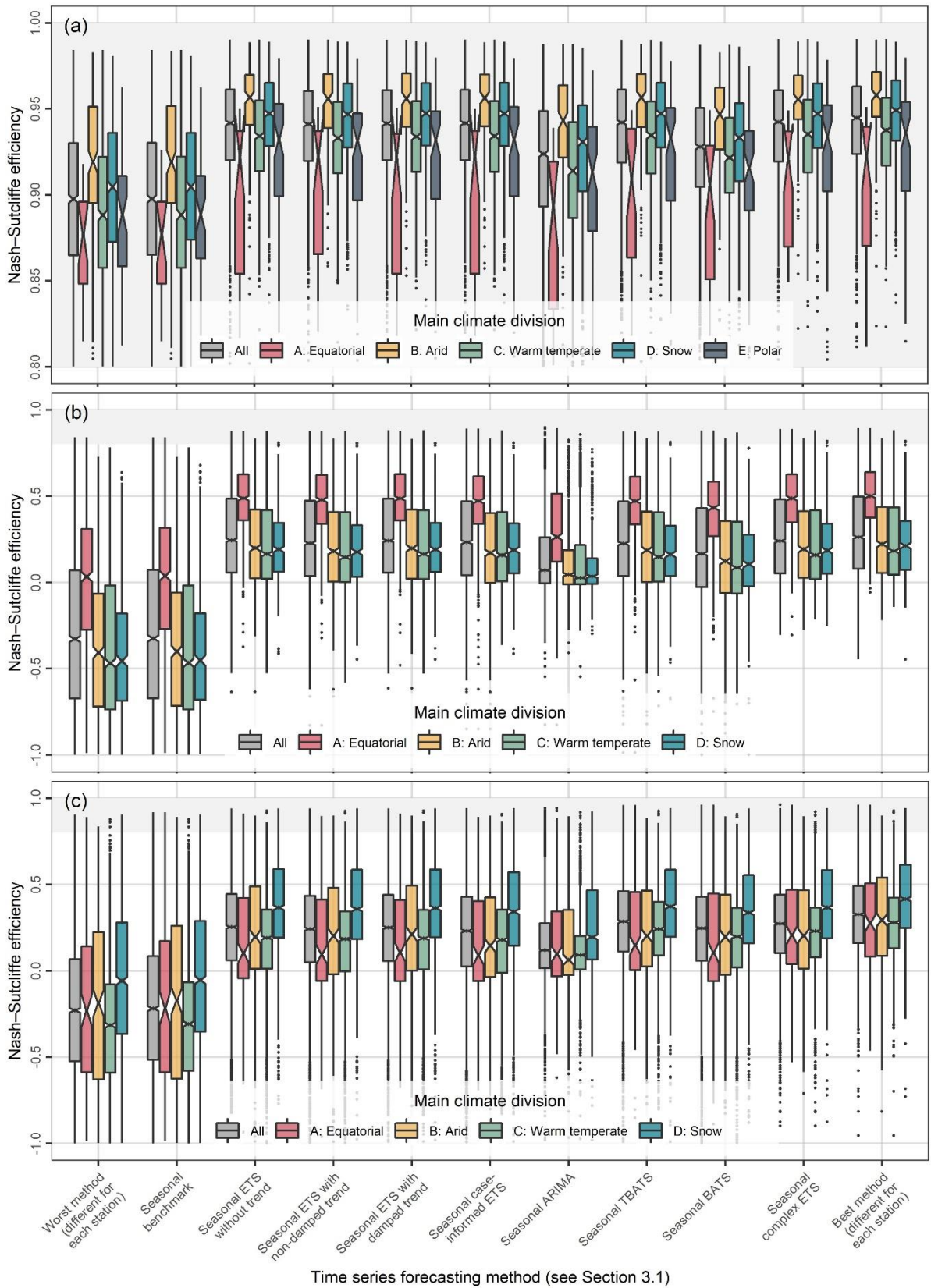

Figure 4. Summary of the results concerning the (a) temperature, (b) precipitation and (c) river flow time series forecastability assessment. The presentation is made per main climate division.

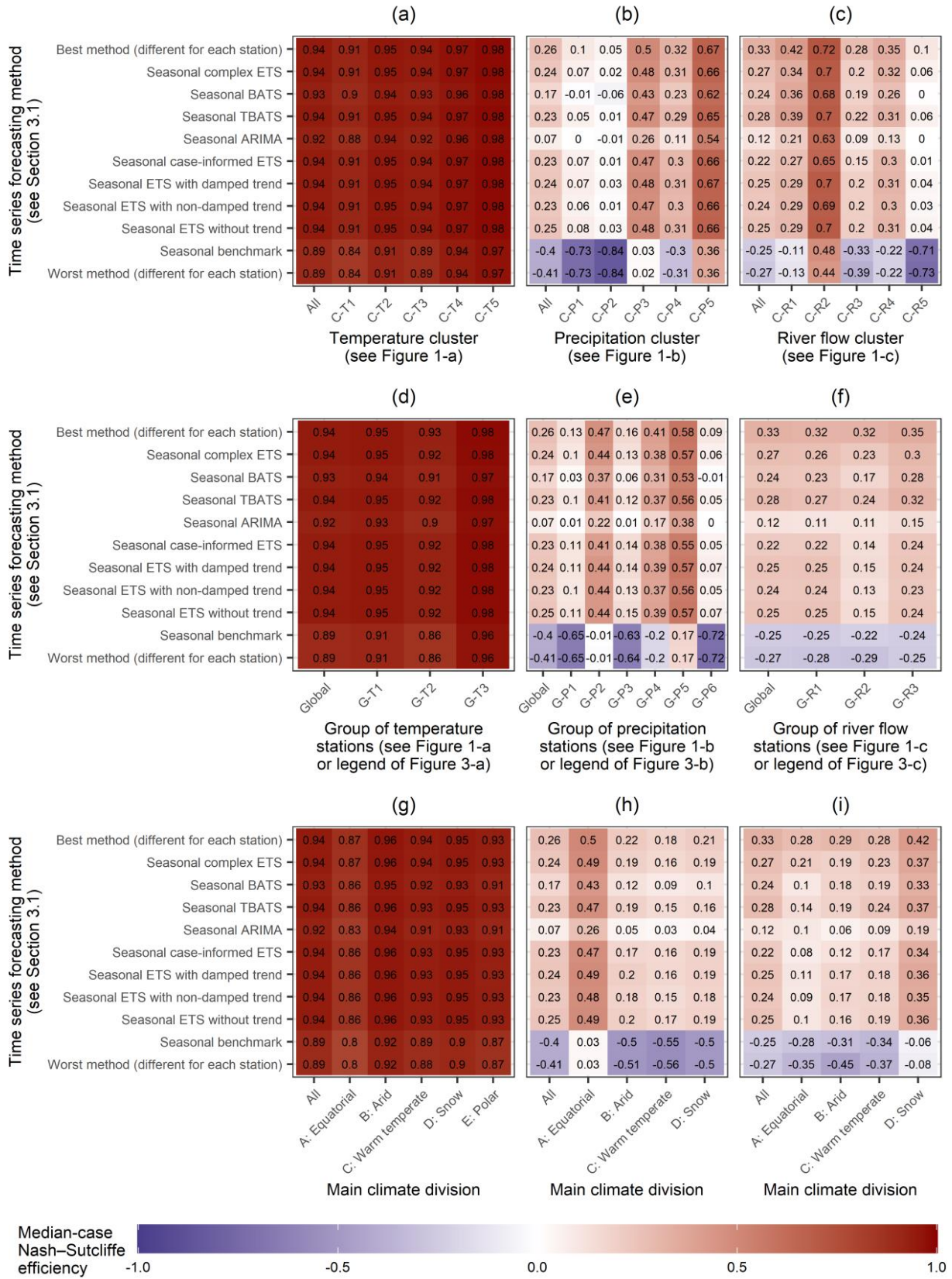

Figure 5. Median-case Nash-Sutcliffe efficiency of the forecasting methods for the (a, d, g) temperature, (b, e, h) precipitation and (c, f, i) river flow time series. The presentation is made per (a–c) cluster, (d–f) group of stations and (g–i) main climate division. The clusters have been obtained using the 40-year-long time series.

## 4.4 Links between time series forecastability and descriptive time series features

We next provide a complete interpretation of 12-month ahead monthly temperature, precipitation and river flow forecastability by quantifying its relationships with the 57 descriptive time series features (see Supplementary Data, Text S3). We also rank these features according to their appropriateness (i.e., the amount of information that they provide) for inferring and foretelling monthly temperature, precipitation and river flow time series forecastability without issuing any forecast for the past.

Figure 6 presents the Spearman correlations computed between the Nash-Sutcliffe efficiency of the forecasts of the best method (which differs for each station) and each of the descriptive features, when the latter have been computed for the entire 40-year-long time series. These specific investigations can effectively reveal the degree of confidence with which we can infer information about the forecastability of a monthly temperature, precipitation or river flow time series by only using descriptive characterizations. They are made separately for each time series type. For all three time series types, the most intense relationships are found for the seasonality strength (`seasonal_strength`; see also the scatterplots in Figure 7j–l) and the sample autocorrelation at lag equal to 12 months (`seas_acf1`). These strong relationships are positive, meaning that the stronger the seasonality features of a time series the larger its forecastability. Time series forecastability is also strongly correlated to the spectral entropy feature (`entropy`) for monthly precipitation and river flow (see also the scatterplots in Figure 7e,f), and the sample entropy feature (`sampen_first`) for monthly temperature and precipitation; however, the respective Spearman correlations are negative, meaning that the larger the `entropy` and `sampen_first` values the smaller the time series forecastability.

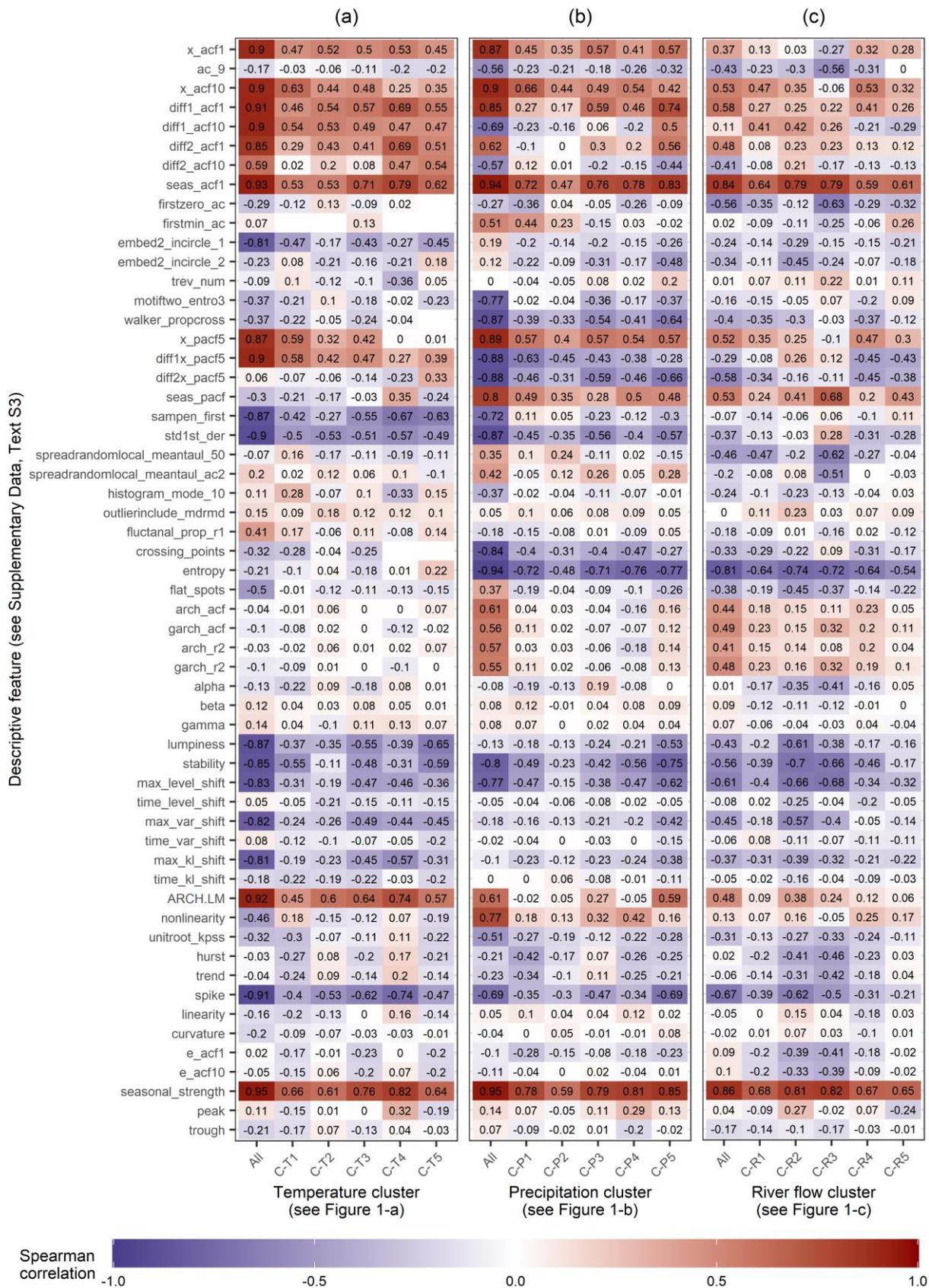

Figure 6. Spearman correlation computed for monthly (a) temperature, (b) precipitation and (c) river flow between the Nash-Sutcliffe efficiency of the best method (which differs for each station) and the values of the descriptive features of the entire 40-year-long time series. Blank cells indicate cases for which the correlation coefficient is not defined.

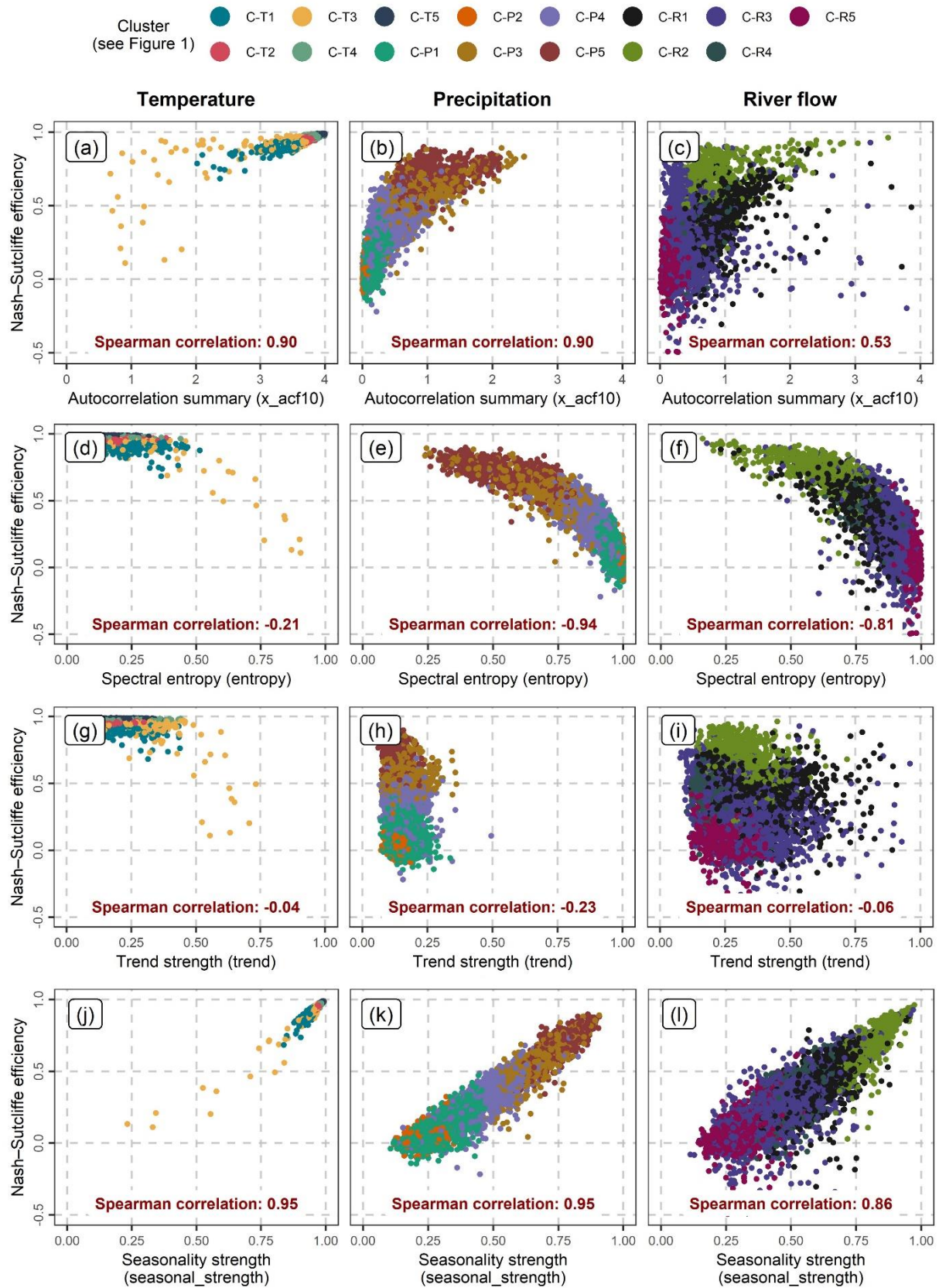

Figure 7. Nash-Sutcliffe efficiency of the best method (which differs for each station) versus descriptive features (a–c) `x_acf10`, (d–f) `entropy`, (g–i) `trend` and (j–l) `seasonal_strength` of the entire 40-year-long monthly (a, d, g, j) temperature, (b, e, h, k) precipitation and (c, f, i, l) river flow time series.

Furthermore, for monthly temperature and precipitation several autocorrelation and partial autocorrelation features (`x_acf1`, `x_acf10`, `diff1_acf1`, `diff1_acf10`, `diff2_acf1`, `x_pacf5`, `diff1x_pacf5`) are found to be strongly correlated, either positively or negatively, with the Nash-Sutcliffe efficiency of the forecasts. For monthly river flow, the respective correlations are of medium magnitude. Other descriptive features that are found to have intense relationships or relationships of medium magnitude with the Nash-Sutcliffe efficiency of the forecasts are the standard deviation of the first-order differenced standardized time series (`std1st_der`), features based on 12-month-long tiled (non-overlapping) windows (`lumpiness`, `stability`), features based on 12-month-long sliding (overlapping) windows (`max_level_shift`, `max_var_shift`, `max_kl_shift`), an autoregressive conditional heteroscedasticity feature (`ARCH.LM`), a nonlinearity feature (`nonlinearity`) and a feature for measuring the spikiness of time series (`spike`).

Moreover, Figure 8 presents the objectively derived rankings of the descriptive features of the entire 40-year-long time series according to their importance in inferring the forecastability of a time series within a non-linear regression setting, and by extension their appropriateness for characterizing a time series with respect to its forecastability without issuing any forecast. Independently of the time series type, the two most informative features for this latter task are `seasonal_strength` and `seas_acf1`, while the third most informative feature is `entropy` for monthly precipitation and river flow, and `max_level_shift` for monthly temperature. Among the top-10 informative descriptive features –at least for one of the examined hydrometeorological time series types– are also the following ones: `x_acf1`, `x_acf10`, `diff1_acf1`, `diff1_acf10`, `x_pacf5`, `diff1x_pacf5`, `diff2x_pacf5`, `sampen_first`, `lumpiness`, `stability`, `max_var_shift`, `ARCH.LM` and `spike`. Other descriptive features found to be among the top-15 informative ones are the following: `ac_9`, `firstzero_ac`, `embed2_incircle_1`, `embed2_incircle_2`, `motiftwo_entro_3`, `walker_propcross`, `std1st_der`, `spreadrandomlocal_meantaul_50`, `crossing_points`, `max_kl_shift`, `nonlinearity` and `linearity`.

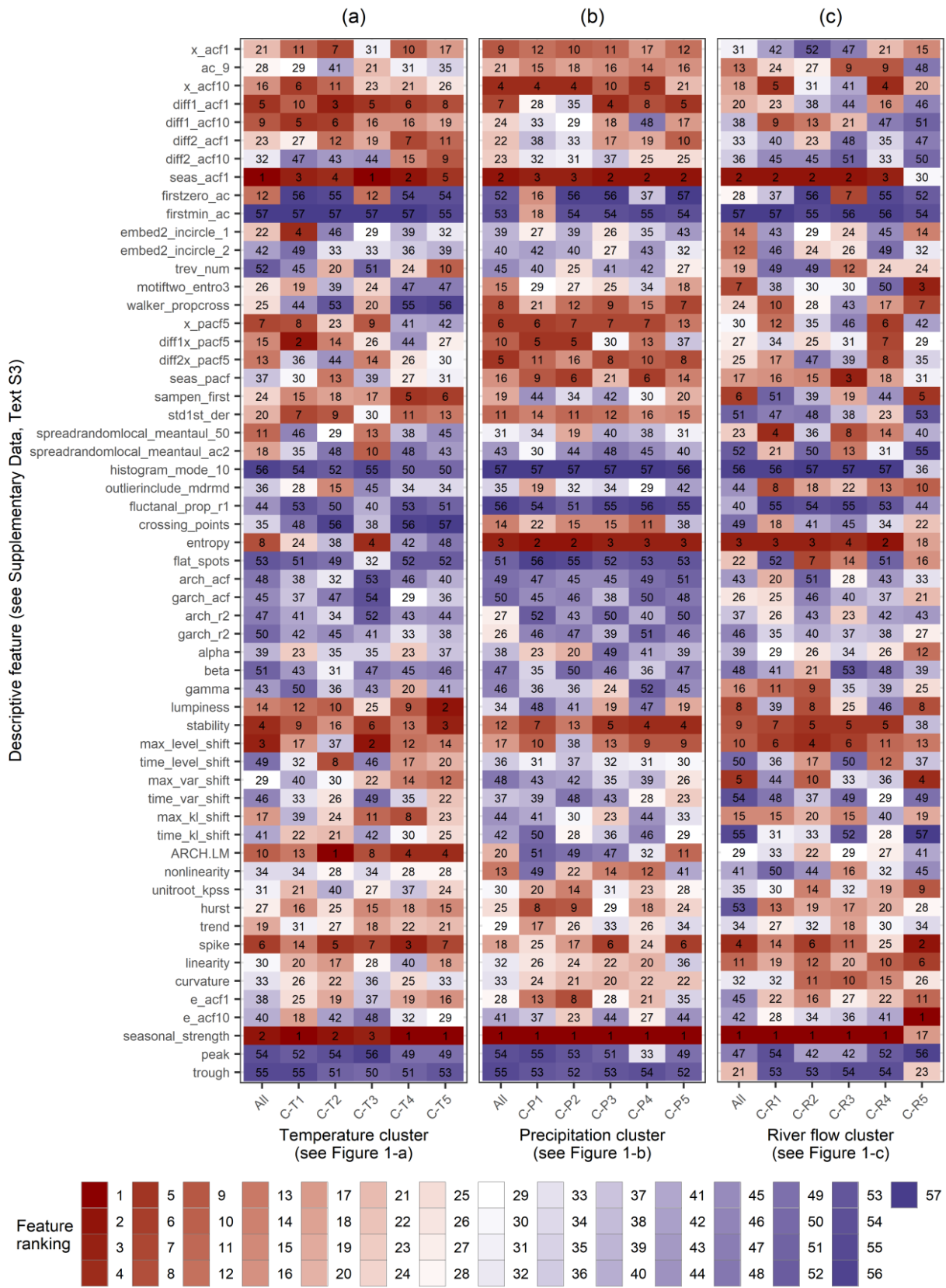

Figure 8. Rankings of the descriptive features of the entire 40-year-long time series according to their relative importance in predicting the Nash-Sutcliffe efficiency of the best method (which differs for each station) for the monthly (a) temperature, (b) precipitation and (c) river flow time series.

The rankings presented in Figure 8 provide a clear view of the most informative features for inferring and interpreting monthly hydroclimatic time series forecastability (with `seasonal_strength`, `seas_acf1` and `entropy` being on the top of the entire list). However, based on Figure 9 we feel that this view could be different, if we had only applied a single time series forecasting method, especially for monthly river flow but also for monthly precipitation (and almost not at all for monthly temperature). For example, within a similar framework using seasonal case-informed ETS only (and assuming that time series forecastability is well-represented by the scores of this time series method's forecasts), the features `diff2_acf10`, `crossing_points` and `time_kl_shift` would be identified as the three most informative ones for inferring and interpreting monthly precipitation time series forecastability (see Figure 9b). Notably, the rankings of the descriptive features according to the information they provide in predicting the Nash-Sutcliffe efficiency of the forecasts of the nine time series forecasting methods strongly differ from the rankings computed for time series forecastability (represented by the best method for each station). For the efficiency of most of the nine methods, `nonlinearity` is the most informative feature, while `seasonal_strength`, `seas_acf1` and `entropy` are far behind in the list. Perhaps the above-summarized information indicates some sort of robustness of the proposed framework. Perhaps it also further highlights some dissimilarities between the three examined time series types, and a larger field of study for the improvement of monthly river flow time series forecasting (and monthly precipitation time series forecasting, yet to a less extent) than for the improvement of monthly temperature time series forecasting. Practical implications stemming from the present work are extensively elaborated in Section 5.3.

Lastly, Supplementary Data (Figure S1) (presenting results for the descriptive features of the eldest 30-year-long segments of the time series) serves as an empirical demonstration that, not only we can infer –to a considerable extent– monthly temperature, precipitation and river flow forecastability in terms of Nash-Sutcliffe efficiency without issuing any forecast for the past, but we can also foretell it (e.g., for the next ten years, as made in our experiment), and all the more with almost the same degree of confidence (see Figure 6), and with the same descriptive features being the most appropriate and informative ones for the task (see Figure 8), as if we were interested in assessments for the past.

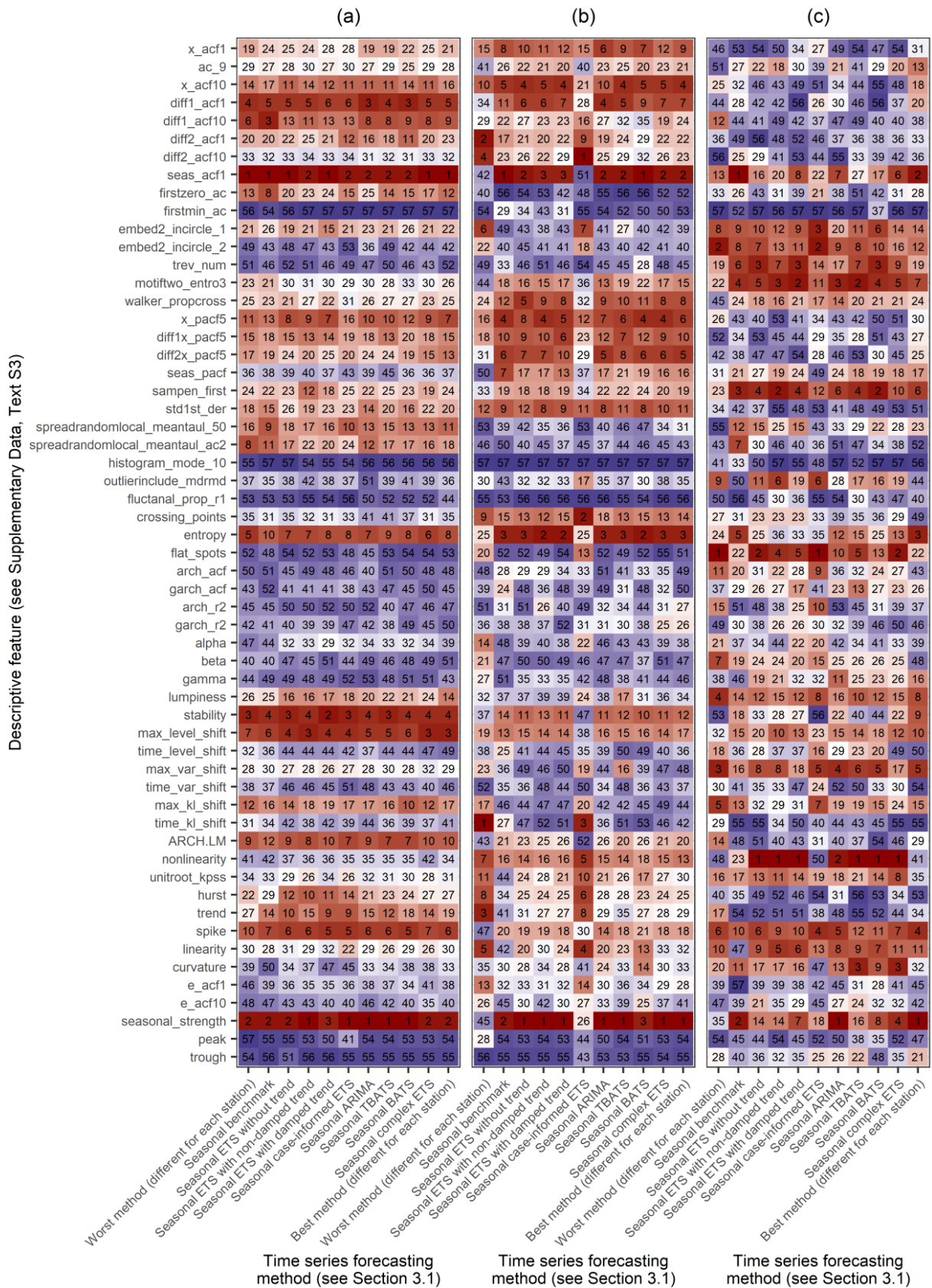

Figure 9. Rankings of the descriptive features of the entire 40-year-long time series according to their relative importance in predicting the Nash-Sutcliffe efficiency offered by the various time series forecasting methods for the monthly (a) temperature, (b) precipitation and (c) river flow time series. For the legend, see Figure 8.

## 5. Discussion

### 5.1 On interpreting monthly hydroclimatic time series forecastability

Among others, the relative importance investigations between 57 largely diverse descriptive time series features (i.e., seasonality, autocorrelation, partial autocorrelation, long-range dependence, entropy, trend, linearity, nonlinearity, stationarity and other features) have offered several results that perhaps would not be anticipated based on the common trends and discussions in the literature (see, e.g., the literature overview in Papacharalampous et al. 2021), and did not confirm results that perhaps would be anticipated and expected (see again Figure 8a,b,c and Supplementary Data, Figure S1b). For instance, several currently under-exploited descriptive features and feature classes (e.g., `spike`, `stability`, `spreadrandomlocal_meantaul_50`, `lumpiness`, `std1st_der`, `flat_spots`, `motiftwo_entro3`, `max_level_shift`, `max_var_shift`, `ARCH.LM`) have been shown to be more important than (or of similar importance to) the various autocorrelation features in interpreting and foretelling monthly hydrometeorological time series forecastability at the global scale (thereby also suggesting the importance of massive feature extraction). Furthermore, and despite possible expectations, the time series trend strength (`trend`), the Hurst parameter of the fractional Gaussian noise process (`hurst`) and a stationarity feature (`unitroot_kpss`) are not among the most informative features for monthly temperature, precipitation and river flow. Perhaps, however, long-range dependence and trends would be mostly expected to have a large impact for forecasting horizons longer than one year.

Moreover, even the dominance of the seasonality and entropy features in interpreting monthly temperature, precipitation and river flow time series forecastability should be further discussed in light of the literature, which perhaps emphasizes more temporal dependence and trends, and especially the latter when the interest lies in future behaviours and changes. Seasonality and entropy features have indeed been found on the top of a long list of descriptive time series features exhibiting either strong or medium-magnitude relationships with time series forecastability. The former features are sometimes perceived in the time series forecasting literature as measures of the "deterministic" component of the time series attributed to seasons (see, e.g., Hyndman and Athanasopoulos 2021, Chapter 3.2), while the latter features are perceived as measures of the "random" or "noisy" component of the time series (see, e.g., Hyndman and Athanasopoulos 2021, Chapter 4.4). In this view, the intense relationships identified for these categories of descriptive features (with some exceptions for the `entropy` and `sampen_first` features computed for monthly temperature and monthly river flow, respectively; see again Figure 8a,b,c and Supplementary Data, Figure S1b) are, in a simplistic sense, in line with the "no free lunch theorem" (Wolpert 1996). In fact, this theorem suggests that, the better (worse) our knowledge of the system to be modelled the better (worse) the modelling solutions can be and, therefore, the larger (smaller) the system's predictability-forecastability is perceived to be (as predictability-forecastability is assessed by using predictive-forecasting models). This is, of course, only a naïve-benchmark interpretation of monthly (or seasonal) time series forecastability and does not mean, by any chance, that the relationships with the remaining features should be forgotten (especially, given that many of them are also intense). In fact, such a simplistic approach would be equivalent to forecasting monthly (or seasonal) time series by using a single seasonal component (obtained through time series decomposition) and expecting the delivered forecasting solutions to be perfect. Furthermore, it has been shown that other descriptive features (e.g., `nonlinearity` and `flat_spots`) can be more informative in predicting the performances of various methods in monthly hydroclimatic time series forecasting (see again Figure 9).

Altogether, it could be stated that monthly temperature, precipitation and river flow time series are characterized more by similarities than by differences as regards the relationships between their forecastability and their descriptive features (e.g., their seasonality, temporal dependence, entropy and trend features). Of course, this does not mean, by any chance, that there are not notable differences between the three examined time series types (see, e.g., the differences in terms of `entropy`, `spike`, `x_pacf5`, `trend` and `lumpiness` in Figures 6 and 8; see also Supplementary Data, Figure S1). Also, it does not mean that their forecastability can be inferred with the same degree of confidence (as, in fact, it cannot with this confidence being notably smaller for river flow). This also suggests the importance of the sample splitting strategy (i.e., the formal way to assess predictive-forecasting models in terms of their generalization ability and to characterize a time series with respect to its actual predictability-forecastability) in obtaining new knowledge. Based on our massive investigations, this importance seems to be more complementary than contrasting to the knowledge obtained through time series analysis.

## 5.2 On hydrometeorological and hydroclimatic time series forecasting at scale

Discussions should also focus on the multiple and diverse benefits stemming from hydrometeorological (and hydroclimatic) time series forecasting at scale by exclusively using time series (otherwise referred to as "stochastic") models (i.e., exponential smoothing, and autoregressive integrated moving average – ARIMA or autoregressive fractionally integrated moving average – ARFIMA models). Although this specific concept has been introduced and exploited before in hydrology by Papacharalampous et al. (2018, 2019), Tyralis et al. (2021) and companion works based on inspirations and blueprints sourced from other scientific fields (e.g., from the works by Makridakis et al. 1987; Chatfield 1988; Makridakis and Hibon 2000; Hyndman and Khandakar 2008; Boulesteix et al. 2018; Taylor and Letham 2018), comprehensive interpretations of its results have been achieved for the first time in this work. On the other hand, this same concept has further allowed us herein to effectively extend –again in a largely interpretable way– the concept of massive feature extraction in water and environmental science (Papacharalampous et al. 2021) from the predictive and predictability-forecastability perspectives. In fact, time series models have been found to be more suitable for such extensions than machine learning algorithms for time series forecasting (although the performance of these two model categories does not differ much in absence of informative exogenous predictors; Papacharalampous et al. 2019). The reasons behind their better suitability are (i) their ability to model temporal dependence, which suggests a clear connection with the field of stochastic hydrology, and (ii) the fact that (most of) these models also offer a certain degree of interpretability. Within large-sample studies, this interpretability allows informative time series feature comparisons from the predictive and predictability-forecastability perspectives (see, e.g., the comparisons between time series models with and without trend parameters in Figures 2–5), thereby complementing well large-sample descriptive time series characterizations and explorations.

## 5.3 On future exploitations and expansions of the proposed framework

Some last discussions should be made on several possible future exploitations and extensions-expansions of our methodological framework. Different temporal scales and processes could be investigated for providing new insights into the interplay between hydrometeorological and hydroclimatic time series analysis and time series forecasting, especially in light of some notable differences identified across the three time series types examined herein (see, e.g., the relevant discussions in Section 5.1). The enrichment of the proposed methodological framework with different mathematical techniques (see, e.g., the ones proposed by Juez and Nadal-Romero 2020, 2021) is also highly recommended. Indeed, such techniques could provide additional (complementary or even contrasting) insights into the problem of interest. Furthermore, more detailed investigations of spatial patterns of time series forecastability would require a special focus and further elaboration of parts of the proposed feature-based framework, such as the inclusion of more clusters (see, e.g., the numbers of clusters selected by Pechlivanidis et al. 2020 and Girons Lopez et al. 2021 for the cases of Europe and Sweden, respectively). Such investigations have been out of the scope of this work, which emphasizes the explanation-interpretation of monthly hydroclimatic time series forecastability through massive relationship investigations.

Moreover, possible extensions could focus on whether (and how) the relative performance of two or more time series forecasting models (assessed by selecting a benchmark and by computing the relative improvements provided by the other methods with respect to this benchmark in terms of any performance score of the forecaster's choice) is related to descriptive time series features. This would require a massive expansion of previous investigations by Papacharalampous and Tyralis (2020), and could be particularly important from a practical point of view, as it could facilitate case-informed integrations of multiple diverse time series forecasting models within systematic cross-learning and feature-based frameworks (see, e.g., Talagala et al. 2018; Montero-Manso et al. 2020). For instance, a forecasting method that is found to significantly outperform others for specific feature values (or specific clusters) could be selected for forecasting other time series with similar features (or time series attributed to the same clusters based on their past), even if it performs badly for other feature values and clusters (and even if it performs badly for most modelling cases and would, therefore, be rejected within a single-method framework). In fact, by extending the computations behind Figure 8 for the nine time series forecasting methods of this work (investigations not presented for reasons of brevity), we have found that even the most important features can differ from cluster to cluster (to a degree rather applying more to monthly river flow than to monthly temperature and precipitation). These observations might indicate that cross-learning and feature-based frameworks could indeed provide notable improvements in hydrological forecasting.

Finally, as regards signature-based time series clustering methodologies for the regionalization of hydrological catchment models (see Pechlivanidis et al. 2020; Girons Lopez et al. 2021), the concept of massive feature extraction (as well as our compilation of 57 descriptive features itself) could be merged with features-signatures that are traditionally of interest in catchment hydrology (see, e.g., the features examined in Iliopoulou et al. 2019; Tyralis et al. 2019b; Knoben et al. 2020; Pechlivanidis et al. 2020; Girons Lopez et al. 2021) for investigating the possibility of providing improvements of the clustering solutions by exploiting additional information contributed by seasonality, entropy and more river flow time series features.

## 6. Conclusions

In this work, we have developed a detailed framework for progressing and deepening our understanding of hydrometeorological and hydroclimatic time series forecastability (and its spatial patterns), and for providing its comprehensive explanation-interpretation through massive feature extraction (and feature-based time series clustering). By conducting three global-scale applications, we have further demonstrated that these specific objectives are largely feasible. These applications are primarily based on nine well-established and/or well-tested time series forecasting methods from the forecasting field, 57 largely diverse descriptive features (i.e., seasonality, autocorrelation, partial autocorrelation, long-range dependence, entropy, trend, linearity, nonlinearity, stationarity and other features), random forests, and over 13 000 monthly temperature, precipitation and river flow time series. As we have examined three different types of monthly hydroclimatic time series, we believe to have provided a broad analysis and overview of monthly hydroclimatic time series forecastability.

The key findings of this work are the following:

o Monthly temperature, precipitation and river flow time series forecastability in terms of Nash-Sutcliffe efficiency can be inferred and foretelled with a rather considerable (even large) degree of confidence by computing descriptive time series features.

o This degree of confidence differs significantly across the three examined time series types. Also, notable differences have been found in the relative importance and the objective rankings of the descriptive time series features according to their contribution in inferring and foretelling forecastability.

- The most informative features to be used in this regard are the strength of seasonality (`seasonal_strength`) and the lag-12 sample autocorrelation (`seas_acf1`), followed by the spectral entropy of the time series (`entropy`) for monthly precipitation and river flow (but not for monthly temperature).

- The above highlighted features have been found on the top of a list containing 57 diverse descriptive features (see Supplementary Data, Text S3), but they are not the only informative ones, and any exclusive focus on them could only lead to a rather naïve-benchmark understanding of the investigated relationships and to naïve-benchmark improvements of our forecasting frameworks.

- In fact, comparably intense relationships or relationships of medium magnitude have been found between time series forecastability and many other descriptive time series features (including features that are under-exploited and mostly new for hydrology), while even the least informative features can find a non-harming place within properly designed machine learning feature-based frameworks (in the sense that there are statistical and machine learning algorithms that automatically perform intrinsic variable selection and others that are not affected by redundancy in their predictors by construction). Details on this and the above key findings can be found in Section 4.4.

- In light of the above points, reasons for higher (lower) monthly hydroclimatic time series forecastability might include, among others, larger (smaller) strength of seasonality, smaller (larger) spectral entropy, larger (smaller) autocorrelation (e.g., at lags 1 and 12), smaller (larger) temporal variation, and smaller (larger) spikiness.

- Distinct spatial patterns of monthly temperature, precipitation and river flow time series forecastability around the globe have emerged and have been thouroughfully interpreted by this work, as detailed in Sections 4.1 and 4.3.

- Comparisons across the main climatic types have also revealed district patterns of monthly temperature, precipitation and river flow time series forecastability. These latter patterns are detailed in Sections 4.2 and 4.3.

Overall, we believe that the massive and automatic character of the proposed framework has been the key to providing the first formal and trustable answers to our research questions herein. This is because such answers could only be provided through extensive and comparative quantifications of the relationships between a variety of descriptive time series features and actual time series forecastability (formally assessed according to the definition of forecasting, which rotates around the unknown future), and such quantifications are missing from the previous literature. This probably holds because of the methodological and computational challenges that such massive quantifications impose. These challenges have only recently started to become manageable through data science; however, they are expected to be easier to tackle in the future, allowing even more massive investigations. We deem that possible extensions-expansions of the proposed framework could have important practical implications (especially if the massive character is retained), as they could, among others, constitute a direct basis for case-informed integrations of diverse time series forecasting methods within systematic frameworks. The high potential of using multiple methods (instead of using a single method) is increasingly recognized in both the time series forecasting and machine learning fields.

**Acknowledgements:** We greatly thank the Editor M. Santosh and the Associate Editor V. Samuel for handling the review process. We also greatly thank the eponymous Referee O. Ledvinka and two anonymous Referees for their constructive and fruitful suggestions, which helped us to substantially improve this work. Funding from the Italian Ministry of Environment, Land and Sea Protection (MATTM) for the SimPRO project (2020–2021) is acknowledged by (in alphabetical order): S. Grimaldi, G. Papacharalampous and E. Volpi. E. Volpi also acknowledges funding from the Italian Ministry of Education, University and Research (MIUR), in the frame of the Departments of Excellence Initiative 2018–2022, attributed to the Department of Engineering of Roma Tre University. I. Pechlivanidis acknowledges funding from the EU Horizon 2020 project CLINT (Climate Intelligence: Extreme events detection, attribution and adaptation design using machine learning) under Grant Agreement 101003876.

**Declarations of interest:** We declare no conflict of interest.

**Author contributions:** GP and HT conceptualized the work, designed the experiments, produced the forecasts, and performed the analyses and visualizations. GP prepared the first draft, which was commented and enriched with interpretations and discussions by HT, IP, SG and EV.

## Appendix A  Data availability

The original hydrometeorological datasets can be retrieved through the following links:

- https://www.ncdc.noaa.gov/ghcnm/v4.php (temperature; Menne et al. 2018)
- https://www.ncdc.noaa.gov/ghcnm/v2.php (precipitation; Peterson and Vose 1997)
- https://doi.org/10.1594/PANGAEA.887477 (river flow; Do et al. 2018)

Gridded climate data (Kottek et al. 2006) can be retrieved through the R package `kgc` (Bryant et al. 2017).

## Appendix B  Supplementary data

Supplementary data can be found in the online version.

## References

Abrahart RJ, See LM, Dawson CW (2008) Neural network hydroinformatics: Maintaining scientific rigour, in: Abrahart RJ, See LM, Solomatine DP (Eds) Practical Hydroinformatics. Springer-Verlag Berlin Heidelberg, pp 33–47. https://doi.org/10.1007/978-3-540-79881-1_3

Addor N, Newman AJ, Mizukami N, Clark MP (2017) The CAMELS data set: Catchment attributes and meteorology for large-sample studies. Hydrology and Earth System Sciences 21:5293–5313. https://doi.org/10.5194/hess-21-5293-2017

Akaike H (1974) A new look at the statistical model identification. IEEE Transactions on Automatic Control 19(6):716–723. https://doi.org/10.1109/TAC.1974.1100705

Alpaydin E (2010) Introduction to machine learning, second edition. MIT Press

Althoff D, Dias SHB, Filgueiras R, Rodrigues LN (2020) ETo-Brazil: A daily gridded reference evapotranspiration data set for Brazil (2000–2018). Water Resources Research 56(7):e2020WR027562. https://doi.org/10.1029/2020WR027562

Althoff D, Rodrigues LN, Bazame HC (2021) Uncertainty quantification for hydrological models based on neural networks: The dropout ensemble. Stochastic Environmental Research and Risk Assessment 35(5):1051–1067. https://doi.org/10.1007/s00477-021-01980-8

Alvarez-Garreton C, Mendoza PA, Boisier JP, Addor N, Galleguillos M, Zambrano-Bigiarini M, Lara A, Puelma C, Cortes G, Garreaud R, McPhee J, Ayala A (2018) The CAMELS-CL dataset: Catchment attributes and meteorology for large sample studies – Chile dataset. Hydrology and Earth System Sciences 22:5817–5846. https://doi.org/10.5194/hess-22-5817-2018

Armstrong JS (2001) Principles of Forecasting. Springer US. https://doi.org/10.1007/978-0-306-47630-3

Beran J (1994) Statistics for Long-Memory Processes. CRC press

Blöschl G, Bierkens MFP, Chambel A, Cudennec C, Destouni G, Fiori A, Kirchner JW, McDonnell JJ, Savenije HHG, Sivapalan M, et al. (2019a) Twenty-three Unsolved Problems in Hydrology (UPH) – A community perspective. Hydrological Sciences Journal 64(10):1141–1158. https://doi.org/10.1080/02626667.2019.1620507

Blöschl G, Hall J, Viglione A, Perdigão RA, Parajka J, Merz B, Lun D, Arheimer B, Aronica GT, Bilibashi A, et al. (2019b) Changing climate both increases and decreases European river floods. Nature 573(7772):108–111. https://doi.org/10.1038/s41586-019-1495-6

Boulesteix A-L, Binder H, Abrahamowicz M, Sauerbrei W (2018) On the necessity and design of studies comparing statistical methods. Biometrical Journal 60(1):216–218. https://doi.org/10.1002/bimj.201700129

Box GEP, Jenkins GM (1970) Time Series Analysis: Forecasting and Control. Holden-Day Inc., San Francisco, USA

Breiman L (2001a) Random forests. Machine Learning 45(1):5–32. https://doi.org/10.1023/A:1010933404324

Breiman L (2001b) Statistical modeling: The two cultures. Statistical Science 16(3):199–231. https://doi.org/10.1214/ss/1009213726

Brown RG (1959) Statistical Forecasting for Inventory Control. McGraw-Hill Book Co., New York, USA

Bryant C, Wheeler NR, Rubel F, French RH (2017) kgc: Köppen-Geiger Climatic Zones. R package version 1.0.0.2. https://CRAN.R-project.org/package=kgc

Ceola S, Laio F, Montanari A (2019) Global-scale human pressure evolution imprints on sustainability of river systems. Hydrology and Earth System Sciences 23(9):3933–3944. https://doi.org/10.5194/hess-23-3933-2019

Chagas VBP, Chaffe PLB, Addor N, Fan FM, Fleischmann AS, Paiva RCD, Siqueira VA (2020) CAMELS-BR: Hydrometeorological time series and landscape attributes for 897 catchments in Brazil. Earth System Science Data 12:2075–2096, https://doi.org/10.5194/essd-12-2075-2020

Chatfield C (1988) What is the 'best' method of forecasting? Journal of Applied Statistics 15(1):19–38. https://doi.org/10.1080/02664768800000003

Cheng KS, Lien YT, Wu YC, Su YF (2017) On the criteria of model performance evaluation for real-time flood forecasting. Stochastic Environmental Research and Risk Assessment 31(5):1123–1146. https://doi.org/10.1007/s00477-016-1322-7

Coxon G, Addor N, Bloomfield JP, Freer J, Fry M, Hannaford J, Howden NJK, Lane R, Lewis M, Robinson EL, Wagener T, Woods R (2020) CAMELS-GB: Hydrometeorological time series and landscape attributes for 671 catchments in Great Britain. Earth System Science Data 12(4):2459–2483. https://doi.org/10.5194/essd-12-2459-2020

Curceac S, Ternynck C, Ouarda TBMJ, Chebana F, Niang SD (2019) Short-term air temperature forecasting using nonparametric functional data analysis and SARMA models. Environmental Modelling and Software 111:394–408. https://doi.org/10.1016/j.envsoft.2018.09.017

Curceac S, Atkinson PM, Milne A, Wu L, Harris P (2020) Adjusting for conditional bias in process model simulations of hydrological extremes: An experiment using the North Wyke Farm Platform. Frontiers in Artificial Intelligence 3:82. https://doi.org/10.3389/frai.2020.565859

De Gooijer JG, Hyndman RJ (2006) 25 years of time series forecasting. International Journal of Forecasting 22(3):443–473. https://doi.org/10.1016/j.ijforecast.2006.01.001

De Livera AM, Hyndman RJ, Snyder RD (2011) Forecasting time series with complex seasonal patterns using exponential smoothing. Journal of the American Statistical Association 106(496):1513–1527. https://doi.org/10.1198/jasa.2011.tm09771

Dimitriadis P, Koutsoyiannis D, Tzouka K (2016) Predictability in dice motion: How does it differ from hydro-meteorological processes? Hydrological Sciences Journal 61(9):1611–1622. https://doi.org/10.1080/02626667.2015.1034128

Dimitriadis P, Koutsoyiannis D, Iliopoulou T, Papanicolaou P (2021) A global-scale investigation of stochastic similarities in marginal distribution and dependence structure of key hydrological-cycle processes. Hydrology 8(2):59. https://doi.org/10.3390/hydrology8020059

Do HX, Gudmundsson L, Leonard M, Westra S (2018) The Global Streamflow Indices and Metadata Archive (GSIM) – Part 1: The production of a daily streamflow archive and metadata. Earth System Science Data 10(2):765–785. https://doi.org/10.5194/essd-10-765-2018

Donoho D (2017) 50 years of data science. Journal of Computational and Graphical Statistics 26(4):745–766. https://doi.org/10.1080/10618600.2017.1384734

Fang Y, Ceola S, Paik K, McGrath G, Rao PSC, Montanari A, Jawitz JW (2018) Globally universal fractal pattern of human settlements in river networks. Earth's Future 6(8):1134–1145. https://doi.org/10.1029/2017EF000746

Fildes R (2020) Learning from forecasting competitions. International Journal of Forecasting 36(1):186–188. https://doi.org/10.1016/j.ijforecast.2019.04.012

Fulcher BD, Jones NS (2017) hctsa: A computational framework for automated time-series phenotyping using massive feature extraction. Cell Systems 5(5):527–531. https://doi.org/10.1016/j.cels.2017.10.001

Fulcher BD, Little MA, Jones NS (2013) Highly comparative time-series analysis: The empirical structure of time series and their methods. Journal of the Royal Society Interface 10(83):20130048. https://doi.org/10.1098/rsif.2013.0048

Fowler KJA, Acharya SC, Addor N, Chou C, Peel MC (2021) CAMELS-AUS: Hydrometeorological time series and landscape attributes for 222 catchments in Australia. Earth System Sciences Data 13, 3847–3867. https://doi.org/10.5194/essd-13-3847-2021

Gardner Jr ES (2006) Exponential smoothing: The state of the art—Part II. International Journal of Forecasting 22(4):637–666. https://doi.org/10.1016/j.ijforecast.2006.03.005

Girons Lopez M, Crochemore L, Pechlivanidis IG (2021) Benchmarking an operational hydrological model for providing seasonal forecasts in Sweden. Hydrology and Earth System Sciences 25:189–1209. https://doi.org/10.5194/hess-25-1189-2021

Goerg GM (2013) Forecastable component analysis. International Conference on Machine Learning, pp. 64–72

Grimaldi S, Kao SC, Castellarin A, Papalexiou SM, Viglione A, Laio F, Aksoy H, Gedikli A (2011) Statistical hydrology, in: Wilderer PA (Ed) Treatise on Water Science 2. Elsevier, pp 479–517

Gudmundsson L, Do HX, Leonard M, Westra S (2018) The Global Streamflow Indices and Metadata Archive (GSIM) – Part 2: Quality control, time-series indices and homogeneity assessment. Earth System Science Data 10:787–804. https://doi.org/10.5194/essd-10-787-2018

Gupta HV, Perrin C, Blöschl G, Montanari A, Kumar R, Clark MP, Andréassian V (2014) Large-sample hydrology: A need to balance depth with breadth. Hydrology and Earth System Sciences 18:463–477. https://doi.org/10.5194/hess-18-463-2014

Harrigan S, Prudhomme C, Parry S, Smith K, Tanguy M (2018) Benchmarking ensemble streamflow prediction skill in the UK. Hydrology and Earth System Sciences 22(3):2023–2039. https://doi.org/10.5194/hess-22-2023-2018

Hastie T, Tibshirani R, Friedman JH (2009) The Elements of Statistical Learning: Data Mining, Inference and Prediction, second edition. Springer, New York. https://doi.org/10.1007/978-0-387-84858-7

Hipel KW, McLeod AI (1994) Time Series Modelling of Water Resources and Environmental Systems, Elsevier. ISBN 978-0-444-89270-6

Holt CC (2004) Forecasting seasonals and trends by exponentially weighted moving averages. International Journal of Forecasting 20(1):5–10. https://doi.org/10.1016/j.ijforecast.2003.09.015

Hyndman RJ (2020) A brief history of forecasting competitions. International Journal of Forecasting 36(1):7–14. https://doi.org/10.1016/j.ijforecast.2019.03.015

Hyndman RJ, Khandakar Y (2008) Automatic time series forecasting: The forecast package for R. Journal of Statistical Software 27(3):1–22. https://doi.org/10.18637/jss.v027.i03

Hyndman RJ, Athanasopoulos G (2021) Forecasting: Principles and Practice. OTexts: Melbourne, Australia. https://otexts.com/fpp3

Hyndman RJ, Koehler AB, Snyder RD, Grose S (2002) A state space framework for automatic forecasting using exponential smoothing methods. International Journal of Forecasting 18(3):439–454. https://doi.org/10.1016/S0169-2070(01)00110-8

Hyndman RJ, Akram M, Archibald BC (2008a) The admissible parameter space for exponential smoothing models. Annals of the Institute of Statistical Mathematics 60(2):407–426. https://doi.org/10.1007/s10463-006-0109-x

Hyndman RJ, Koehler AB, Ord JK, Snyder RD (2008b) Forecasting with exponential smoothing: The state space approach. Springer-Verlag Berlin Heidelberg, pp 3–7. https://doi.org/10.1007/978-3-540-71918-2

Hyndman RJ, Wang E, Laptev N (2015) Large-scale unusual time series detection. 2015 IEEE International Conference on Data Mining Workshop (ICDMW), Atlantic City, NJ, pp 1616–1619. https://doi.org/10.1109/ICDMW.2015.104

Hyndman RJ, Kang Y, Montero-Manso P, Talagala T, Wang E, Yang Y, O'Hara-Wild M (2020) tsfeatures: Time Series Feature Extraction. R package version 1.0.2. https://CRAN.R-project.org/package=tsfeatures

Iliopoulou T, Aguilar C, Arheimer B, Bermúdez M, Bezak N, Ficchì A, Koutsoyiannis D, Parajka J, Polo MJ, Thirel G, Montanari A (2019) A large sample analysis of European rivers on seasonal river flow correlation and its physical drivers. Hydrology and Earth System Sciences 23(1):73–91. https://doi.org/10.5194/hess-23-73-2019

James G, Witten D, Hastie T, Tibshirani R (2013) An Introduction to Statistical Learning. Springer, New York. https://doi.org/10.1007/978-1-4614-7138-7

Juez C, Nadal-Romero E (2020) Long-term time-scale bonds between discharge regime and catchment specific landscape traits in the Spanish Pyrenees. Environmental Research 191:110158. https://doi.org/10.1016/j.envres.2020.110158

Juez C, Nadal-Romero E (2021) Long-term temporal structure of catchment sediment response to precipitation in a humid mountain badland area. Journal of Hydrology 597:125723. https://doi.org/10.1016/j.jhydrol.2020.125723

Kang Y, Hyndman RJ, Smith-Miles K (2017) Visualising forecasting algorithm performance using time series instance spaces. International Journal of Forecasting 33(2):345–358. https://doi.org/10.1016/j.ijforecast.2016.09.004

Kang Y, Hyndman RJ, Li F (2020) GRATIS: GeneRAting TIme Series with diverse and controllable characteristics. Statistical Analysis and Data Mining: The ASA Data Science Journal 13(4):354–376. https://doi.org/10.1002/sam.11461

Khatami S, Peel MC, Peterson TJ, Western AW (2019) Equifinality and flux mapping: A new approach to model evaluation and process representation under uncertainty. Water Resources Research 55(11):8922–8941. https://doi.org/10.1029/2018WR023750

Khatami S, Peterson TJ, Peel MC, Western AW (2020) Evaluating catchment models as multiple working hypotheses: On the role of error metrics, parameter sampling, model structure, and data information content. https://doi.org/10.1002/essoar.10504066.1

Klingler C, Schulz K, Herrnegger M (2021) LamaH-CE: LArge-SaMple DAta for Hydrology and Environmental Sciences for Central Europe. Earth Syst. Sci. Data 13:4529–4565. https://doi.org/10.5194/essd-13-4529-2021

Knoben WJM, Freer JE, Peel MC, Fowler KJA, Woods RA (2020) A brief analysis of conceptual model structure uncertainty using 36 models and 559 catchments. Water Resources Research 56(9):e2019WR025975. https://doi.org/10.1029/2019WR025975

Kottek M, Grieser J, Beck C, Rudolf B, Rubel F (2006) World map of the Köppen-Geiger climate classification updated. Meteorologische Zeitschrift 15(3):259–263. https://doi.org/10.1127/0941-2948/2006/0130

Koutsoyiannis D (2010) HESS Opinions "A random walk on water". Hydrology and Earth System Sciences 14:585–601. https://doi.org/10.5194/hess-14-585-2010

Koutsoyiannis D (2013) Hydrology and change. Hydrological Sciences Journal 58(6):1177–1197. https://doi.org/10.1080/02626667.2013.804626

Ledvinka O (2015) Evolution of low flows in Czechia revisited. Proceedings of the International Association of Hydrological Sciences 369:87–95. https://doi.org/10.5194/piahs-369-87-2015

Ledvinka O, Lamacova A (2015) Detection of field significant long-term monotonic trends in spring yields. Stochastic Environmental Research and Risk Assessment 29(5):1463–1484. https://doi.org/10.1007/s00477-014-0969-1

Makridakis S, Hibon M, Lusk E, Belhadjali M (1987) Confidence intervals: An empirical investigation of the series in the M-competition. International Journal of Forecasting 3(3–4):489–508. https://doi.org/10.1016/0169-2070(87)90045-8

Makridakis S, Hibon M (2000) The M3-Competition: Results, conclusions and implications. International Journal of Forecasting 16(4):451–476. https://doi.org/10.1016/S0169-2070(00)00057-1

Manero Font J, Béjar Alonso J (2021) Forecastability measures that describe the complexity of a site for deep learning wind predictions. Supercomputing Frontiers and Innovations, 8(1):8–27. https://doi.org/10.14529/jsfi210102

Mandelbrot BB, Wallis JR (1968) Noah, Joseph, and operational hydrology. Water Resources Research 4(5):909–918. https://doi.org/10.1029/WR004i005p00909

Markonis Y, Koutsoyiannis D (2013) Climatic variability over time scales spanning nine orders of magnitude: Connecting Milankovitch cycles with Hurst–Kolmogorov dynamics. Surveys in Geophysics 34(2):181–207. https://doi.org/10.1007/s10712-012-9208-9

Markonis Y, Hanel M, Máca P, Kyselý J, Cook ER (2018a) Persistent multi-scale fluctuations shift European hydroclimate to its millennial boundaries. Nature Communications 9(1):1–12. https://doi.org/10.1038/s41467-018-04207-7

Markonis Y, Moustakis Y, Nasika C, Sychova P, Dimitriadis P, Hanel M, Máca P, Papalexiou SM (2018b) Global estimation of long-term persistence in annual river runoff. Advances in Water Resources 113:1–12. https://doi.org/10.1016/j.advwatres.2018.01.003

Menne MJ, Williams CN, Gleason BE, Rennie JJ, Lawrimore JH (2018) The global historical climatology network monthly temperature dataset, version 4. Journal of Climate 31(24):9835–9854. https://doi.org/10.1175/JCLI-D-18-0094.1

Messager ML, Lehner B, Cockburn C, Lamouroux N, Pella H, Snelder T, Tockner K, Trautmann T, Watt C, Datry T (2021) Global prevalence of non-perennial rivers and streams. Nature 594:391–397. https://doi.org/10.1038/s41586-021-03565-5

Moallemi EA, de Haan FJ, Hadjikakou M, Khatami S, Malekpour S, Smajgl A, Stafford Smith M, Voinov A, Bandari R, Lamichhane P, Miller KK, et al. (2021) Evaluating participatory modeling methods for co-creating pathways to sustainability. Earth's Future 9(3):e2020EF001843. https://doi.org/10.1029/2020EF001843

Montanari A (2003) Long-range dependence in hydrology, in: Doukhan P, Oppenheim G, Taqqu M (Eds) Theory and Applications of Long-Range Dependence. Boston: Birkhauser, pp. 461–472

Montanari A, Young G, Savenije HHG, Hughes D, Wagener T, Ren LL, Koutsoyiannis D, Cudennec C, Toth E, Grimaldi S, et al. (2013) "Panta Rhei—Everything Flows": Change in hydrology and society—The IAHS Scientific Decade 2013–2022. Hydrological Sciences Journal 58(6):1256–1275. https://doi.org/10.1080/02626667.2013.809088

Montero-Manso P, Athanasopoulos G, Hyndman RJ, Talagala TS (2020) FFORMA: Feature-based forecast model averaging. International Journal of Forecasting 36(1):86–92. https://doi.org/10.1016/j.ijforecast.2019.02.011

Nash JE, Sutcliffe JV (1970) River flow forecasting through conceptual models part I—A discussion of principles. Journal of Hydrology 10(3):282–290. https://doi.org/10.1016/0022-1694(70)90255-6

Newman AJ, Clark MP, Sampson K, Wood A, Hay LE, Bock A, Viger RJ, Blodgett D, Brekke L, Arnold JR, Hopson T, Duan Q (2015) Development of a large-sample watershed-scale hydrometeorological data set for the contiguous USA: Data set characteristics and assessment of regional variability in hydrologic model performance. Hydrology and Earth System Sciences 19:209–223. https://doi.org/10.5194/hess-19-209-2015

Papacharalampous GA, Tyralis H (2020) Hydrological time series forecasting using simple combinations: Big data testing and investigations on one-year ahead river flow predictability. Journal of Hydrology 590:125205. https://doi.org/10.1016/j.jhydrol.2020.125205

Papacharalampous GA, Tyralis H, Koutsoyiannis D (2018) Predictability of monthly temperature and precipitation using automatic time series forecasting methods. Acta Geophysica 66(4):807–831. https://doi.org/10.1007/s11600-018-0120-7

Papacharalampous GA, Tyralis H, Koutsoyiannis D (2019) Comparison of stochastic and machine learning methods for multi-step ahead forecasting of hydrological processes. Stochastic Environmental Research and Risk Assessment 33(2):481–514. https://doi.org/10.1007/s00477-018-1638-6

Papacharalampous GA, Tyralis H, Papalexiou SM, Langousis A, Khatami S, Volpi E, Grimaldi S (2021) Global-scale massive feature extraction from monthly hydroclimatic time series: Statistical characterizations, spatial patterns and hydrological similarity. Science of the Total Environment 767:144612. https://doi.org/10.1016/j.scitotenv.2020.144612

Papalexiou SM, Montanari A (2019) Global and regional increase of precipitation extremes under global warming. Water Resources Research 55(6):4901–4914. https://doi.org/10.1029/2018WR024067

Pechlivanidis IG, Arheimer B (2015) Large-scale hydrological modelling by using modified PUB recommendations: The India-HYPE case. Hydrology and Earth System Sciences 19:4559–4579. https://doi.org/10.5194/hess-19-4559-2015

Pechlivanidis IG, Crochemore L, Rosberg J, Bosshard T (2020) What are the key drivers controlling the quality of seasonal streamflow forecasts? Water Resources Research 56(6):e2019WR026987. https://doi.org/10.1029/2019WR026987

Peterson TC, Vose RS (1997) An overview of the Global Historical Climatology Network Temperature database. Bulletin of the American Meteorological Society 78(12):2837–2850. https://doi.org/10.1175/1520-0477(1997)078<2837:AOOTGH>2.0.CO;2

Ponce-Flores M, Frausto-Solís J, Santamaría-Bonfil G, Pérez-Ortega J, González-Barbosa JJ (2020) Time series complexities and their relationship to forecasting performance. Entropy 22(1):89. https://doi.org/10.3390/e22010089

Python Software Foundation (2021) Python Language Reference. http://www.python.org

Quilty J, Adamowski J (2020) A stochastic wavelet-based data-driven framework for forecasting uncertain multiscale hydrological and water resources processes. Environmental Modelling and Software 130:104718. https://doi.org/10.1016/j.envsoft.2020.104718

Quilty J, Adamowski J, Boucher MA (2019) A stochastic data-driven ensemble forecasting framework for water resources: A case study using ensemble members derived from a database of deterministic wavelet-based models. Water Resources Research 55(1):175–202. https://doi.org/10.1029/2018WR023205

R Core Team (2021) R: A language and environment for statistical computing. R Foundation for Statistical Computing, Vienna, Austria. https://www.R-project.org

Rahman ATMS, Hosono T, Kisi O, Dennis B, Imon AHMR (2020) A minimalistic approach for evapotranspiration estimation using the Prophet model. Hydrological Sciences Journal 65(12):1994–2006. https://doi.org/10.1080/02626667.2020.1787416

Scheidegger AE (1970) Stochastic models in hydrology. Water Resources Research 6(3):750–755. https://doi.org/10.1029/WR006i003p00750

Serinaldi F, Chebana F, Kilsby CG (2020) Dissecting innovative trend analysis. Stochastic Environmental Research and Risk Assessment 34:733–754. https://doi.org/10.1007/s00477-020-01797-x

Shmueli G (2010) To explain or to predict? Statistical Science 25(3):289–310. https://doi.org/10.1214/10-STS330

Sikorska-Senoner AE, Seibert J (2020) Flood-type trend analysis for alpine catchments. Hydrological Sciences Journal 65(8):1281–1299. https://doi.org/10.1080/02626667.2020.1749761

Sikorska-Senoner AE, Quilty JM (2021) A novel ensemble-based conceptual-data-driven approach for improved streamflow simulations. Environmental Modelling and Software 143:105094. https://doi.org/10.1016/j.envsoft.2021.105094

Sivakumar B, Berndtsson R (2010) Advances in Data-Based Approaches for Hydrologic Modeling and Forecasting. World Scientific Publishing Company, Singapore. https://doi.org/10.1142/7783

Sidle RC (2021) Strategies for smarter catchment hydrology models: Incorporating scaling and better process representation. Geoscience Letters 8:24. https://doi.org/10.1186/s40562-021-00193-9
Spearman C (1904) The proof and measurement of association between two things. The American Journal of Psychology 15(1):72–101. https://doi.org/10.2307/1412159
Svetunkov I, Kourentzes N (2016) Complex Exponential Smoothing. Lancaster University Management School, Lancaster, pp 1–31
Széles B, Broer M, Parajka J, Hogan P, Eder A, Strauss P, Blöschl G (2018) Separation of scales in transpiration effects on low flows: A spatial analysis in the Hydrological Open Air Laboratory. Water Resources Research 54(9):6168–6188. https://doi.org/10.1029/2017WR022037
Széles B, Parajka J, Hogan P, Silasari R, Pavlin L, Strauss P, Blöschl G (2021) Stepwise prediction of runoff using proxy data in a small agricultural catchment. Journal of Hydrology and Hydromechanics 69(1):65–75. https://doi.org/10.2478/johh-2020-0029
Taieb SB, Bontempi G, Atiya AF, Sorjamaa A (2012) A review and comparison of strategies for multi-step ahead time series forecasting based on the NN5 forecasting competition. Expert Systems with Applications 39(8):7067–7083. https://doi.org/10.1016/j.eswa.2012.01.039
Talagala TS, Hyndman RJ, Athanasopoulos G (2018) Meta-learning how to forecast time series. Working Paper 6/18. Department of Econometrics and Business Statistics, Monash University
Talagala TS, Li F, Kang Y (2019) FFORMPP: Feature-based forecast model performance prediction. https://arxiv.org/abs/1908.11500
Taylor JW (2003) Exponential smoothing with a damped multiplicative trend. International Journal of Forecasting 19(4):715-725. https://doi.org/10.1016/S0169-2070(03)00003-7
Taylor SJ, Letham B (2018) Forecasting at scale. The American Statistician 72(1):37–45. https://doi.org/10.1080/00031305.2017.1380080
Todini E (2007) Hydrological catchment modelling: Past, present and future. Hydrology and Earth System Sciences 11:468–482. https://doi.org/10.5194/hess-11-468-2007
Tyralis H, Papacharalampous GA (2021) Boosting algorithms in energy research: A systematic review. Neural Computing and Applications 33, 14101–14117. https://doi.org/10.1007/s00521-021-05995-8
Tyralis H, Papacharalampous GA, Langousis A (2019a) A brief review of random forests for water scientists and practitioners and their recent history in water resources. Water 11(5):910. https://doi.org/10.3390/w11050910
Tyralis H, Papacharalampous GA, Tantanee S (2019b) How to explain and predict the shape parameter of the generalized extreme value distribution of streamflow extremes using a big dataset. Journal of Hydrology 574:628–645. https://doi.org/10.1016/j.jhydrol.2019.04.070
Tyralis H, Papacharalampous GA, Langousis A (2021) Super ensemble learning for daily streamflow forecasting: Large-scale demonstration and comparison with multiple machine learning algorithms. Neural Computing and Applications 33:3053–3068. https://doi.org/10.1007/s00521-020-05172-3
Volpi E (2019) On return period and probability of failure in hydrology. Wiley Interdisciplinary Reviews: Water 6(3):e1340. https://doi.org/10.1002/wat2.1340
Wei WWS (2006) Time Series Analysis, Univariate and Multivariate Methods, second edition. Pearson Addison Wesley

Winters PR (1960) Forecasting sales by exponentially weighted moving averages. Management of Forecasting 6(3):324–342. https://doi.org/10.1287/mnsc.6.3.324

Wolpert DH (1996) The lack of a priori distinctions between learning algorithms. Neural Computation 8(7):1341–1390. https://doi.org/10.1162/neco.1996.8.7.1341

Xu L, Chen N, Zhang X, Chen Z (2018) An evaluation of statistical, NMME and hybrid models for drought prediction in China. Journal of Hydrology 566:235–249. https://doi.org/10.1016/j.jhydrol.2018.09.020

Yan D, Chen A, Jordan MI (2013) Cluster forests. Computational Statistics and Data Analysis 66:178–192. https://doi.org/10.1016/j.csda.2013.04.010

Yevjevich VM (1987) Stochastic models in hydrology. Stochastic Hydrology and Hydraulics 1(1):17–36. https://doi.org/10.1007/BF01543907